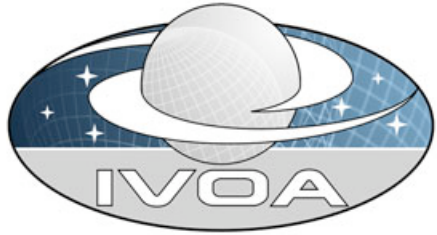

# VOResource: an XML Encoding Schema for Resource Metadata
# Version 1.03

## IVOA Recommendation
## 22 February 2008

**Authors:**
Raymond Plante, Editor
Kevin Benson
Matthew Graham
Gretchen Greene
Paul Harrison
Gerard Lemson
Tony Linde
Guy Rixon
Aurélien Stébé
and the IVOA Registry Working Group.

## Abstract

This document describes an XML encoding standard for IVOA Resource Metadata, referred to as VOResource. This schema is primarily intended to support interoperable registries used for discovering resources; however, any application that needs to describe resources may use this schema. In this document, we define the types and elements that make up the schema as representations of metadata terms defined in the IVOA standard, Resource Metadata for the Virtual Observatory [Hanisch et al. 2004]. We also describe the general model for the schema and explain how it may be extended to add new metadata terms and describe more specific types of resources.

## Status of this document

This document has been produced by the IVOA Resource Registry Working Group.
It has been reviewed by IVOA Members and other interested parties, and has been endorsed by the IVOA Executive Committee as an IVOA Recommendation as of 2007 September 27. It is a stable document and may be used as reference material or cited as a normative reference from another document. IVOA's role in making the Recommendation is to draw attention to the specification and to promote its widespread deployment. This enhances the functionality and interoperability inside the Astronomical Community.

A list of current IVOA Recommendations and other technical documents can be found at http://www.ivoa.net/Documents/.

# Acknowledgements

This document has been developed with support from the National Science Foundation's Information Technology Research Program under Cooperative Agreement AST0122449 with The Johns Hopkins University, from the UK Particle Physics and Astronomy Research Council (PPARC), and from the Eurpean Commission's Sixth Framework Program via the Optical Infrared Coordination Network (OPTICON).

## Conformance-related definitions

The words "MUST", "SHALL", "SHOULD", "MAY", "RECOMMENDED", and "OPTIONAL" (in upper or lower case) used in this document are to be interpreted as described in IETF standard, RFC 2119 [RFC 2119].

The **Virtual Observatory (VO)** is general term for a collection of federated resources that can be used to conduct astronomical research, education, and outreach. The **International Virtual Observatory Alliance (IVOA)** is a global collaboration of separately funded projects to develop standards and infrastructure that enable VO applications.

XML document **validation** is a software process that checks that an XML document is not only well-formed XML but also conforms to the syntax rules defined by the applicable schema. Typically, when the schema is defined by one or more XML Schema [Schema] documents (see next section), validation refers to checking for conformance to the syntax described in those Schema documents. This document describes additional syntax constraints that cannot be enforced solely by the rules of XML Schema; thus, in this document, use of the term validation includes the extra checks that goes beyond common Schema-aware parsers which ensure conformance with this document.

**UTC** refers to *Universal Coordinated Time* as defined by....

## Syntax Notation Using XML Schema

The eXtensible Markup Language, or XML, is document syntax for marking textual information with named tags and is defined by the World Wide Web Consortium (W3C) Recommendation, XML 1.0 [XML]. The set of XML tag names and the syntax rules for their use is referred to as the document schema. One way to formally define a schema for XML documents is using the W3C standard known as XML Schema [Schema].

This document defines the VOResource schema using XML Schema. The full Schema document is listed in Appendix A. Parts of the schema appear within the main sections of this document; however, documentation nodes have been left out for the sake of brevity.

Reference to specific elements and types defined in the VOResource schema include the namespaces prefix, `vr`, as in `vr:Resource` (a type defined in the VOResource schema). Use of the `vr` prefix in compliant instance documents is not required (see section 2.1); its use in this document is simply to indicate that it is an entity defined in the VOResource schema.

# Contents

# 1. Introduction

The IVOA Standard, Resource Metadata for the Virtual Observatory [Hanisch et al. 2004] (hereafter referred to as **RM**) defines metadata terms for describing resources. The RM defines a resource as:

> ... VO element that can be described in terms of who curates or maintains it and which can be given a name and a unique identifier. Just about anything can be a resource: it can be an abstract idea, such as sky coverage or an instrumental setup, or it can be fairly concrete, like an organization or a data collection. This definition is consistent with its use in the general Web community as "anything that has an identity" (Berners-Lee 1998, IETF RFC2396). We expand on this definition by saying that it is also describable.

The resource metadata are, then, the terms and concepts that describe a resource in general. The RM defines the terms as well as describes reasonable or allowed values; it does not, however, describe how the terms and values should be encoded. This is because resource metadata may be encoded in several different formats, depending on the context. This document specifically describes an encoding called VOResource.

The primary intended use of VOResource is to provide an XML interchange format for use with resource registries. A registry is a repository of resource descriptions [RM] and is employed by users and applications to discover resources. VOResource can be used to pass descriptions from publishers to registries and then from registries to applications. Another inended use is as a language for services to describe themselves directly. VOResource may be used in other ways, in whole or in part, using the standard XML mechanisms (e.g., import, include).

The VOResource schema provides XML encoding for so-called core metadata from the RM that (with a few exceptions) can apply to all resources; however, it is recognized that a full and useful description of a *specific* resource will require additional metadata that is relevant only to a resource of its type. Thus, the VOResource schema has been especially designed to be extended. The model for doing this is described in section XX.

---

**Note:**
> The name "VOResource" has in practice had two meanings within IVOA discussions. The first refers specifically to the core XML schema defined by this document. The second refers more broadly to the core schema plus the set of legal extensions. In this document, use of the name "VOResource" corresponds to the first meaning.
> Reference to the broader set of schemas will be indicated explicitly with the annotating phrase, "and its legal extensions."

---

# 2. The VOResource Data Model

The primary use for VOResource, of course, is to describe a resource using the metadata concepts defined in the RM. Here's an example of a VOResource document describing an organisation, the Radio Astronomy Imaging Group at the National Center for Supercomputing Applications.

---

**Example**

A description of an organisation as a resource, organisation.xml

```
   <?xml version="1.0" encoding="UTF-8"?>
2  <resource xsi:type="Organisation"
1    xmlns:vr="http://www.ivoa.net/xml/VOResource/v1.0"
     xmlns:xsi="http://www.w3.org/2001/XMLSchema-instance"
```

```
3    xsi:schemaLocation="http://www.ivoa.net/xml/VOResource/v1.0
3                         http://www.ivoa.net/xml/VOResource/v1.0">

4      <title>NCSA Radio Astronomy Imaging</title>
4      <shortName>NCSA-RAI</shortName>
4      <identifier>ivo://rai.ncsa/RAI</identifier>
4
4      <curation>
5        <publisher ivo-id="ivo://ncsa.uiuc/NCSA">
5          National Center for Supercomputing Applications
5        </publisher>
6        <creator>
6          <name>   Dr. Richard Crutcher    </name>
6          <logo>
6             http://rai.ncsa.uiuc.edu/rai.jpg
6          </logo>
6        </creator>
4        <date>1993-01-01</date>
4        <contact>
4          <name>Dr. Raymond Plante</name>
4          <email>rplante@ncsa.uiuc.edu</email>
4        </contact>
4      </curation>
4
4      <content>
4        <subject>radio astronomy</subject>
4        <subject>data repositories</subject>
4        <subject>digital libraries</subject>
4        <subject>grid-based processing</subject>
4        <description>
4          The Radio Astronomy Imaging Group at the National Center for
4          Supercomputing Applications is focused on applying
4          high-performance computing to astronomical research.  Our
4          projects include the NCSA Astronomy Digital Image Library,
4          the BIMA Data Archive, the BIMA Image Pipeline, and the
4          National Virtual Observatory.
4        </description>
4        <referenceURL>http://rai.ncsa.uiuc.edu/</referenceURL>
4        <type>Organisation</type>
4        <contentLevel>Research</contentLevel>
4      </content>

7      <facility>Berkeley-Illinois-Maryland Array (BIMA)</facility>
7      <facility>
7         Combined Array for Research in Millimeter Astronomy (CARMA)
7      </facility>

    </resource>
```

This example illustrates some important components of a VOResource record:

1. the VOResource namespace,
2. the specific type of resource indicated by the value of the `xsi:type` attribute,
3. the location of the schema documents used by this description,
4. values for the three main types core metadata: identity, curation, and content,
5. a reference to another resource is made by providing that resource's IVOA identifier,
6. string values can be padded with spaces for easier readability,
7. extension metadata specific to the type of resource.

## 2.1. The Schema Namespace and Location

The VOResource schema namespace is "http://www.ivoa.net/xml/VOResource/v1.0". The namespace URI has been chosen to allow it to be resolved as a URL to the XML Schema document (given in Appendix A) that defines the VOResource schema. Applications may assume that the namespace URI is so resolvable.

Authors of instance documents that use the VOResource schema may choose to provide a location for VOResource XML Schema document using the `xsi:schemaLocation` attribute; the choice of the location value is the choice of the author. In general, the use of `xsi:schemaLocation` is recommended by this specification with a the namespace URI given as the location as illustrated in the example above, unless the application prefers otherwise.

```
xsi:schemaLocation="http://www.ivoa.net/xml/VOResource/v1.0
                    http://www.ivoa.net/xml/VOResource/v1.0"
```

Whenever instance validation is needed, use of the VOResource schema and its legal extensions must be declared using the standard namespace declaration attribute, **xmlns:*prefix*** (where *prefix* is an arbitrary prefix). The prefix, **vr**, is used by convention as the prefix defined for the VOResource schema; however, instance documents may use any prefix of the author's choosing. In this document, the **vr** prefix is used to label, as shorthand, a type or element name that is defined in the VOResource schema, as in **vr:Resource**.

Because the VOResource XML schema sets **elementFormDefault="unqualified"**, documents that use the VOResource schema should not use the namespace declaration attribute, **xmlns** (used to set the default namespace), anywhere in the document where the VOResource schema is in effect. (This is a restriction set by the rules of XML Schema.) Furthermore, in accordance with the Schema rules for unqualified elements, the VOResource namespace prefix must not used to qualify VOResource elements. In general, namespace prefixes are only used to qualify type names given as values to the **xsi:type** attribute (see next section). Legal extensions of the VOResource schema SHOULD also set **elementFormDefault="unqualified"** for consistancy.

## 2.2. The Core Structural and Semantic Model

The VOResource schema only defines global types; it does not define any global elements (often refered to as root elements). It is the responsibility of the application to define the root element of the VOResource-employing documents it operates on. Typically, the root element is defined in a separate application-specific schema. The type of an application document's root element is not assumed to be any particular type defined in the VOResource schema (nor any of its legal extensions). In fact, it need not be of a type from the VOResource at all; rather, VOResource types may appear anywhere in the document.

> **Note:**
> The IVOA Registry Interface standard, for example, includes a small schema that defines a single global element, **<VOResources>**, that can contain a series of **<resource>** child elements. The child element is defined to be of the type **Resource** from the VOResource schema.

> **Note:**
> In the example instance document at the beginning of section 2, the root element, **<resource>** is not defined in any schema. Nevertheless, this document is still legal and verifiable XML. This is because the element's type is explicitly specified with the **xsi:type** attribute.

VOResource uses the following conventions for names of elements and types:

- all global types it defines have names that are capitalized. (This practice would extend to global elements, if they existed in the VOResource.)
- Locally defined elements begin with a lower-case character.
- For all types and element names that are made up of multiple words, such as **shortName**, upper-case letters are used demarcate the start of appended word (the "camel" format).
- Names that include abbreviations, such as **IdentifierURI**, all letters in the abbreviation are capitalized.

It is recommended that this convention be followed in other schemas that either use the VOResource schema (via an **xsd:import** or **xsd:include**) or extend it.

Applications describe a single resource using an element of the type **vr:Resource** or a legal derivation of it. The content of the **vr:Resource** type is referred to as the **core VOResource metadata**, and they fall into four categories (corresponding to the sections 3.1, 3.2, 3.3, and 4 of the RM):

- identity metadata: the `<title>`, `<shortName>`, and `<identifier>` elements;
- curation metadata: the contents of the `<curation>` element;
- general content metadata: the contents of the `<content>` element;
- metadata quality flags: the `<validationLevel>` element.

These elements are defined in more detail in section 3.

Many of the elements in VOResource that are meant to have string or URI values are defined as being of the type `xs:token`. This allows authors of VOResource instance documents to pad string and URI values with spaces and include carriage returns to improve readability. The definition of these types will cause an XML Schema-compliant parser to replace tab, line feed, and carriage return characters with simple spaces, then replace multiple sequential occurrences of spaces with a single space, and then remove all leading and trailing spaces.

All VOResource elements and attributes that contain dates must be interpreted as UTC dates and must be encoded in compliance with ISO8601 standard Date and Time Format [ISO8601]. The `vr:UTCTimestamp` type provides a special restriction of the format that requires includsion of date and time, but disallows the timezone format. This enforces a restricted form of this format which allows encoding of the date and time, but disallows the timezone format:

---

**vr:UTCTimestamp Schema Definition**

```
<xs:simpleType name="UTCTimestamp">
    <xs:restriction base="xs:dateTime">
        <xs:pattern value="\d{4}-\d\d-\d\dT\d\d:\d\d:\d\d(\.\d+)?"/>
    </xs:restriction>
</xs:simpleType>
```

**vr:UTCDateTime Schema Definition**

```
<xs:simpleType name="UTCDateTime">
    <xs:union memberTypes="xs:date vr:UTCTimestamp"/>
</xs:simpleType>
```

---

The `vr:UTCDateTime` type allows the date to appear either as a simple date--i.e. conforming to `xs:date`--or as a date and time as restricted by `vr:UTCTimestamp`. When either of these VOResource types is used, authors must provide an applicable UTC date and time, and applications must interpret the time and date as UTC.

All VOResource types and elements have an associated semantic meaning which is given in the first `<xsd:documentation>` node within the type or element's definition in the schema. The meaning associated with a type is generic, indicating the kind of information the type provides. The semantics that are delivered by a VOResource instance document, however, are those associated with VOResource elements. The meaning of a VOResource element can be thought of as having two parts: the generic meaning of the set of information it contains as defined by its type, and a specific meaning describing the context in which that information applies. Because all VOResource elements are locally defined (in the XML Schema sense), they do not have an absolute meaning, but rather have a meaning tied to the thing being described by that element as represented by the enclosing type.

Here are three examples that illustrate the semantics communicated by VOResource entities:

1. The `vr:Curation` type describes the curation of a resource. The `<curation>` element describes curation of the specific resource described by the enclosing `vr:Resource` type and identified by its `identifier` element.

2. The `vr:ResourceName` type is a generic reference to another resource. The `<publisher>` element gives a reference to the publisher of the specific resource being described which may itself be a registered resource described elsewhere.

3. The `<title>` element gives the title of the resource being described the enclosing `vr:Resource` type and identified by its `<identifier>` element. The `<title>` element's type, `xs:token` (a restriction on `xsd:string`), has no inherent meaning associated with it.

Additional semantics are transmitted through the use of derived types using the `xsi:type` attribute. In the sample instance document above, the use of `xsi:type="Organisation"` means that the resource being described is specifically an organisation as defined by the `vr:Organisation` type. This type provides additional metadata that are not part of the core resource metadata. The semantics associated with the use of `xsi:type` is described further in the [next subsection](#).

### 2.2.1. Refined Semantics with Derived Types

When a resource is described with an element explicitly of the type `vr:Resource`, it is being described in the most generic sense. The metadata presented in this type, including both free text values and controlled vocabulary, can give some sense of what type of resource is being described and what it might be used for. However, the most useful descriptions of resources will not explicitly use the `vr:Resource` type; rather, they will use types that are derived from `vr:Resource`.

Defining derived `vr:Resource` types accomplishes two things:

1. it sharpens the semantic meaning of the resource description by indicating what specific type of resource it is, and
2. it *may* allow additional metadata not part of the core but specific to the that type of resource.

The VOResource schema defines two types derived from `vr:Resource`: `vr:Organisation` and `vr:Service`. The `vr:Organisation` adds metadata describing the astronomical facilities such as telescopes that are associated with the organisation it describes. The `vr:Service` type adds an element called `capability` which describes the service's interface as well as information regarding its behavior.

Extensions of the `vr:Resource` type is a key way derivation is used in VOResource to provide refined resource descriptions. Two other important parent types in the schema are `vr:Capability` and `vr:Interface`; these are extended to provide more refined descriptions of services (see [section 2.2.2](#)). The motivation for extending these types are the same as for `vr:Resource`: to provide more specific semantic meaning through the derived type's name, and to provide additional, specialized metadata that is not part of the parent type. Note, however, that in general, a derived type need not alter the content model of its base type. This allows derived types to add more specific meaning with out adding any additional metadata.

As described in [section 2.2](#), it is intended that derived `vr:Resource`, `vr:Capability` and `vr:Interface` types be invoked in instance documents using the `xsi:type` attribute (as illustrated in the [sample document](#) above). This method illustrates a polymorphism for resource metadata in that any place where an element of parent type is expected, the derived type may be inserted. The use of `xsi:type` illustrates both what specific type is being inserted as well as what it is being inserted for. That is, as in our [example](#), the *resource* being described is an *organisation*.

The other mechanism for polymorphism provided by XML Schema [[Schema](#)] is substitution groups. Invoking derived `vr:Resource` types via elements in a substitution group is discouraged because it is less obvious from looking at the instance document that a substitution is being made.

### 2.2.2 The Service Data Model

The `vr:Service` type extends the core `vr:Resource` metadata data by adding the `capability` element (see [section 3.2.2](#)). This element is used to describe a major functionality of the service, usually accessible through a single service endpoint URL. In particular, it is used to describe support for an IVOA service standard (e.g. Simple Image Access Protocol). A service resource record may have multiple child `capability` elements, each describing a different major functionality; however, these capabilities should be related in an obvious, logical way by virtue of sharing same [core VOResource metadata](#).

> **Note:**
> Whether multiple related capabilities are grouped together in a single Service record or are described in separate Service records is expected to be the choice of the VOResource record author. However, it is also expected that resource registry providers will provide some guidance to authors on best practices. This guidance could in part come in the form of a GUI that naturally encourage or contrains to aggregation of capabilities in a single record.

The `capability` element, through its type `vr:Capability`, describes the behavior of service capability and how to access it. The latter is described by a child `interface` element. As for the behavior, the base `vr:Capability` type only provides a `description` element that can contain human-readable text on what this capability provides. More structured behavioral information must be provided through specialized `vr:Capability` extensions. In particular, it is expected that a service standard (e.g. Simple Image Access Protocol) would define an extension of `vr:Capability` that adds additional metadata that can describe the service's behavior in relation to the standard; for example, the added metadata can describe limitations of the service implementation or indicate support for optional features. The specific `vr:Capability` type is invoked using the `xsi:type` mechanism described in section 2.2.1.

---

**Example**

Invoking a specialized Capability type for a standard service capability. In this example, it is assumed that `SimpleImageAccess` extension type is defined in a separate schema document.

```
<capability xsi:type="sia:SimpleImageAccess"
            standardID="ivo://ivoa.net/std/SIA">
  ...
</capability>
```

---

As the example above suggests, a common way for locating or otherwise identifying support for a standard service capability is by looking for the appropriate value in a `capability`'s `xsi:type` attribute.

If the service capability being described does not conform to any standard or if the standard does not require any specialized capability metadata for describing an implementation's behavior, then no `vr:Capability` extension is required. In this case, the base `capability` element is simply used without any `xsi:type` attribute provided. It is expected that the metadata provided outside of the `capability` element as well as (optionally) it's `description` element is sufficient for describing what the service does.

Because each `vr:Capability` extension features a different set of behavioral metadata, introducing a new `vr:Capability` extension can impose a non-trivial cost on applications that process VOResource records. Thus, an alternative way to indicate support for a service standard is provided by the `standardID` attribute which is useful when the standard does not require any specialized behavioral metadata to be provided. The value is set to a URI which represents the service standard. Some service standards that do extend `vr:Capability` *may* force the value of this attribute to be set to the appropriate value (see section 2.3.2); this allows one to use, when appropriate, the `standardID` as a way to locate support for a standard regardless of whether an extension type has been defined or not.

Each `capability` element can contain one or more child `interface` elements, each describing how the capability can be accessed. The `interface` element's type, `vr:Interface`, is abstract; thus, the `interface` element must be accompanied by an `xsi:type` attribute that indicates an `vr:Interface` extension type. The VOResource schema defines two `vr:Interface` extension types: `vr:WebBrowser`, for describing an interface access via web browser, and `vr:WebService`, for accessing a service described by a Web Service Description Language (WSDL) document (see section XX for details).

---

**Example**

Describing a typical SOAP-based Web Service interface.

```
<interface xsi:type="vr:WebService">
  <accessURL> http://archive.org/service/query </accessURL>
</capability>
```

---

When a `capability` contains more than one `interface`, each `interface` should be interpreted as an alternative interface for accessing essentially the same underlying capability. The interfaces can differ in their overall type (e.g. `vr:WebBrowser`, `vr:WebService`) or in the supported input parameters or output products.

When a standard capability is being described (i.e. either the `vr:Capability` sub-type is defined by a standard or the `standardID` is provided), then at least one of the `interface` elements should describe an interface required by the standard. The `role` attribute is used to mark the standard interfaces (typically with the value "std"; see section XX for details). All other interfaces are considered non-standard alternatives.

Another important way `interface`s inside the same `capability` element can be different is in the version of the service standard the interface supports. Whenever an interface supports a version other than "1.0", the `interface` element must include a `version` attribute set to the version being supported. Valid values for `version` are defined by the standard.

<div style="border:1px solid #000; padding:10px;">

**Example**

Describing multiple interfaces for the same capability. In this example, it is assumed that `SkyNode` extension type is defined in a separate schema document.

```
<capability xsi:type="sn:SkyNode"
            standardID="ivo://ivoa.net/std/SkyNode">

  <!--  version 1.0 of the standard SkyNode interface -->
  <interface xsi:type="vr:WebService" role="std" version="1.0">
     <accessURL> http://archive.org/service/skynode </accessURL>
  </interface>

  <!--  version 1.1 of the standard SkyNode interface -->
  <interface xsi:type="vr:WebService" role="std" version="1.1">
     <accessURL> http://archive.org/service/skynode.1.1 </accessURL>
  </interface>

  <!--  a interactive alternative interface, assesible via a browser  -->
  <interface xsi:type="vr:WebBrowser">
     <accessURL> http://archive.org/skynode.html </accessURL>
  </interface>
  ...
</capability>
```

</div>

## 2.3. Extending the VOResource Schema

A schema made up only of global type definitions provides great flexibility for extension. Any global type defined in the VOResource schema may be used as the base of a derived type defined in another schema. The schema containing the derived types must declare its own namespace URI or default to the null namespace; it must not adopt the VOResource namespace URI. The application must then define what schemas, identified by their namespace URIs, are supported and/or required.

A **VOResource extension** is an XML Schema document whose primary purpose is to define new types derived from those defined in the VOResource schema for use in resource descriptions. It is recommended that VOResource extensions follow the definition styles used by the core VOResource. In particular:

- *Provide semantic definitions of all types and elements within the first `<xsd:documentation>` element inside the type or element definition.* Subsequent `<xsd:documentation>` elements may provide additional comments or discussion.

- *Avoid the use of `xsd:choice` elements.* VOResource does not use the choice structure because it does not map readily into any object-oriented software language structure. Choices are handled instead as multiple derived types that can be inserted in place of a parent type.

- *Avoid the use of substitution groups.* VOResource prefers instead the use of `xsi:type` which are (with a few exceptions) functionally equivalent to substitution groups in terms of structure; however, `xsi:type` serves as an obvious flag in the instance document that a substitution has been made.

- *Choose semantically meaningful names for derived types.* When the derived type appears in the pattern `<elname xsi:type="derivedType">`, choose a *derivedType* name such that the sentence, "a *derivedType* is a kind of *elname*" makes natural and obvious sense. For example, "an *Organisation* is a kind of *resource*."

- *Follow the VOResource naming conventions.*

There are two types of derivation that are particularly important to the VOResource data model: derivation of the `vr:Resource` type, used to define specific types of resources, and the derivation of service metadata elements.

### 2.3.1. Defining New Resource Types

Derivation of `vr:Resource` to define new kinds of resources should be done by extension (i.e. using `<xsd:extension>`) rather than restriction. It is not required that the derived type change the content model from that of the `vr:Resource` base type; in this case, the purpose of the derivation is only to sharpen the semantic meaning of the resource description.

### 2.3.2. Defining New Service Capabilities and Interfaces

As described in section 2.2.3, a service standard will often define a new `vr:Capability` extension type to allow implementations to describe how they support the standard. This definition of the `vr:Capability` extension should be done in a schema document with a namespace identifier that is dedicated to that standard (hereafter referred to as *the standard's extension schema*). The extension type should include elements representing the applicable Capability metadata described in section 5.2 of the RM (e.g. *Service.MaxReturnRecords, Service.MaxReturnSize*) but can also include other concepts that are specific to that standard.

The standard's extension schema *may* create a derived `vr:Capability` type that forces the value of the `standardID` attribute to be set to a given URI. This should be done by first deriving from `vr:Capability` by *restriction* (i.e. using `<xsd:restriction>`), keeping all of the parent's content model except adding to the `standardID`'s attribute definition `use="required"` along with the `fixed` modifier set to the desired URI. Since this restricted type is not intended for direct use in an instance document, it should be marked as abstract. The restricted type should then be extended to add the specialized capability metadata required by the standard. (See the example below.)

It is not recommended that standard's extension schema attempt to force the inclusion of a required interface type.

An extension schema can define new interface types, though not necessarily in the context of any specific standard service capability. The basic `vr:Interface` type provides only `accessURL` and `securityMethod` as child elements. A derived `vr:Interface` type must indicate in the documentation how the `<accessURL>` should be interpreted and used. The derived type may also include other added metadata describing how to use the service (e.g., a description of the input arguments). If the interface extension type is expected to be referenced by a standard service capability, then it is recommended that the additional metadata be optional unless the metadata is absolutely required by clients in order to invoke the service.

> **Note:**
> It is intended that a set of common generic interface types would be defined in a separate VOResource extension schema. At the time of this writing, this schema is called VODataService. It currently defines an interface type for describing traditional GET and POST services. More specific interfaces, particularly those associated with standard IVOA services (like a Registry Service) would derive its specific interface descriptions from one of the common types as appropriate.

> **Note:**
> The Simple Image Access Protocol [SIA] is an example of a standard service that defines capability metadata. These include "maxRecords" that list the maximum number of records the SIA implementation can return at a time, and "MaxImageSize" gives the maximum image size that the service can return.

> **Example**
> How a Capability element can be defined to describe the Simple Image Access Protocol...
>
> ```
> <?xml version="1.0" encoding="UTF-8"?>
> <xs:schema xmlns:xs="http://www.w3.org/2001/XMLSchema"
>            xmlns:vr="http://www.ivoa.net/xml/VOResource/v1.0"
>            xmlns:vs="http://www.ivoa.net/xml/VODataService/v1.0"
>            xmlns:sia="http://www.ivoa.net/xml/SIA/v1.0"
>            targetNamespace="http://www.ivoa.net/xml/SIA/v1.0"
>            elementFormDefault="unqualified" attributeFormDefault="unqualified"
>            version="1.0">
>
>     <xs:annotation>
>         <xs:documentation>
>             An XML Schema for describing a Simple Image Access Service
>             implementation.
> ```

```
                </xs:documentation>
        </xs:annotation>

        <xs:import namespace="http://www.ivoa.net/xml/VOResource/v1.0"/>

        <xs:complexType name="SIACapRestriction" abstract="true">
            <xs:annotation>
                <xs:documentation>
                    an abstract capability that fixes the standardID to the
                    URI for the SIA standard.
                </xs:documentation>
            </xs:annotation>
            <xs:complexContent>
                <xs:restriction base="vr:Capability">
                    <xs:sequence>
                        <xs:element name="validationLevel" type="vr:Validation"
                                    minOccurs="0" maxOccurs="unbounded"/>
                        <xs:element name="description" type="vr:token"
                                    minOccurs="0"/>
                        <xs:element name="interface" type="vr:Interface"
                                    minOccurs="0" maxOccurs="unbounded"/>
                    </xs:sequence>
                    <xs:attribute name="standardID" type="vr:IdentifierURI"
                                  use="required" fixed="ivo://ivoa.net/std/SIA"/>
                </xs:restriction>
            </xs:complexContent>
        </xs:complexType>

        <xs:complexType name="SimpleImageAccess">
            <xs:annotation>
                <xs:documentation>
                    The behavior and limitations of an SIA implementation.
                </xs:documentation>
            </xs:annotation>

            <xs:complexContent>
                <xs:extension base="sia:SIACapRestriction">
                    <xs:sequence>

                        <xs:element name="maxRecords" type="xs:int">
                            <xs:annotation>
                                <xs:documentation>
                                    The largest number of records that the Image Query web
                                    method will return.
                                </xs:documentation>
                            </xs:annotation>
                        </xs:element>

                        <!-- other capability metadata -->
                    </xs:sequence>
                </xs:extension>
            </xs:complexContent>
        </xs:complexType>

</xs:schema>
```

...and what a supporting instance document would look like.

```
<resource xsi:type="vr:Service"
    xmlns:vr="http://www.ivoa.net/xml/VOResource/v1.0"
    xmlns:sia="http://www.ivoa.net/xml/SIA/v1.0"
    xmlns:xsi="http://www.w3.org/2001/XMLSchema-instance">

    <!-- the core VOResource metadata -->

    <capability xsi:type="sia:SimpleImageAccess">
        <interface xsi:type="...">
            <!-- interface description -->
            <accessURL> http://archive.org/services/sia <accessURL>
        </interface>

        <maxRecords>5000</maxRecords>
        <!-- other capability metadata -->
    </sia:capability>
```

```
    </resource>
```

# 3. The VOResource Metadata

This section enumerates the types and elements defined in the VOResource schema and describes their meaning in terms of the RM.

## 3.1. The Base Resource Type

A resource, as defined by the RM, is any entity or component of a VO application that is describable and identifiable by a IVOA Identifier. A resource is described using VOResource by an element of the type `vr:Resource` or one of its legal extensions. The schema definition (below) includes elements that encode the identity, curation, and general content metadata for a resource (see sections 3.1 thru 3.3 of the RM). The RM states that certain metadata are required in a minimally compliant resource description; this requirement is enforced by the VOResource schema definition.

---

**vr:Resource Type Schema Definition**

```
<xs:complexType name="Resource">
    <xs:sequence>

        <xs:element name="validationLevel" type="vr:Validation"
                minOccurs="0" maxOccurs="unbounded"/>
        <xs:element name="title" type="vr:token"/>
        <xs:element name="shortName" type="vr:ShortName" minOccurs="0"/>
        <xs:element name="identifier" type="vr:IdentifierURI"/>
        <xs:element name="curation" type="vr:Curation"/>
        <xs:element name="content" type="vr:Content"/>

    </xs:sequence>

    <xs:attribute name="created" type="vr:UTCDateTime"/>
    <xs:attribute name="updated" type="vr:UTCDateTime"/>
    <xs:attribute name="status" default="active">
        <xs:simpleType>
            <xs:restriction base="xs:string">
                <xs:enumeration value="active"/>
                <xs:enumeration value="inactive"/>
                <xs:enumeration value="deleted"/>
            </xs:restriction>
        </xs:simpleType>
    </xs:attribute>
</xs:complexType>
```

---

The child elements for `vr:Resource` are described in subsequent sections.

The `vr:Resource` attributes represent a special class of metadata: they describe the resource metadata description contained within the `vr:Resource` itself as opposed to the resource being described. Their meaning are as follows:

| vr:Resource Attributes | | |
|---|---|---|
| **Attribute** | **Definition** | |
| `created` | *Value type:* | UTC date stamp: `vr:UTCDateTime` |
| | *Semantic Meaning:* | The date this resource metadata description was created. |
| `updated` | *Value type:* | UTC date stamp: `vr:UTCDateTime` |
| | *Semantic Meaning:* | The date this resource metadata description was last updated. |
| `status` | *Value type:* | string, controlled vocabulary: `xsd:string` |
| | *Semantic Meaning:* | a tag indicating whether this resource is believed to be still actively maintained. |
| | *Allowed Values:* | `active` resource is believed to be currently maintained, and its description is up to date (default). |
| | | `inactive` resource is apparently not being maintained at the present. |
| | | `deleted` resource publisher has explicitly deleted the resource. |

The following sections define the elements that encode the specific metadata from the RM. In all cases, the semantic meaning of an element is defined by the RM metadatum it corresponds to (labeled "*RM*

*Name*" below). All rules and advice given in the "Comments" portions in the RM definition apply. Any textual differences in the semantic definitions given below from those given in the RM are intended only for clarification for the XML encoding context.

### 3.1.1. Identity Metadata

The identity metadata described in the RM (section 3.1) are represented as top-level children of the `vr:Resource` type.

| vr:Resource Identity Metadata Elements | |
|---|---|
| **Element** | **Definition** |
| title | *RM Name:* Title<br>*Value type:* string: `xs:token`<br>*Semantic Meaning:* the full name given to the resource<br>*Occurrences:* required |
| shortName | *RM Name:* ShortName<br>*Value type:* string limited to 16 characters or fewer: `vr:ShortName`<br>*Semantic Meaning:* a short name or abbreviation given to the resource.<br>*Occurrences:* optional |
| identifier | *RM Name:* Identifier<br>*Value type:* IVOA identifier URI: `vr:IdentifierURI`<br>*Semantic Meaning:* an unambiguous reference to the resource conforming to the IVOA standard for identifiers [ID]<br>*Occurrences:* required |

Two special types, `vr:ShortName` and `vr:identifierURI` are defined to support identity metadata. The `vr:ShortName` definition enforces the 16-character limit on shortNames.

| vr:ShortName Type Schema Definition |
|---|
| <pre><xs:simpleType name="ShortName"><br>  <xs:restriction base="xs:string"><br>    <xs:maxLength value="16"/><br>  </xs:restriction><br></xs:simpleType></pre> |

The `vr:IdentifierURI` enforces the URI syntax for IVOA Identifiers as defined by the IVOA Identifier standard [ID].

| vr:IdentifierURI Type Schema Definition |
|---|
| <pre><xs:simpleType name="IdentifierURI"><br>  <xs:restriction base="xs:anyURI"><br>    <xs:pattern value="ivo://[\w\d]([\w\d\-_\.!~*'()\+=]{2,}<br>(/[\w\d\-_\.!~*'()\+=]+(/[\w\d\-_\.!~*'()\+=]+)*)?"/><br>  </xs:restriction><br></xs:simpleType></pre> |

Two additional types which are not used within the `vr:Resource` type but are available to support the two components of an IVOA Identifier [ID]: `vr:AuthorityID` and `vr:ResourceKey`.

| vr:AuthorityID Type Schema Definition |
|---|
| <pre><xs:simpleType name="AuthorityID"><br>  <xs:restriction base="xs:string"><br>    <xs:pattern value="[\w\d][\w\d\-_\.!~*'()\+=]{2,}"/><br>  </xs:restriction><br></xs:simpleType></pre> |

| vr:ResourceKey Type Schema Definition |
|---|
| <pre><xs:simpleType name="ResourceKey"><br>  <xs:restriction base="xs:string"><br>    <xs:pattern value="[\w\d\-_\.!~*'()\+=]+(/[\w\d\-_\.!~*'()\+=]+)*"/></pre> |

```
        </xs:restriction>
    </xs:simpleType>
```

### 3.1.2. Curation Metadata

The curation metadata described in the RM (section 3.2) are bundled together into the `vr:Resource` child element, `<curation>`. Its content is defined by the `vr:Curation` complex type.

---

**vr:Curation Type Schema Definition**

```
<xs:complexType name="Curation">
  <xs:sequence>

    <xs:element name="publisher" type="vr:ResourceName"/>
    <xs:element name="creator" type="vr:Creator"
        minOccurs="0" maxOccurs="unbounded"/>
    <xs:element name="contributor" type="vr:ResourceName"
        minOccurs="0" maxOccurs="unbounded"/>
    <xs:element name="date" type="vr:Date"
        minOccurs="0" maxOccurs="unbounded"/>
    <xs:element name="version" type="xs:token" minOccurs="0"/>
    <xs:element name="contact" type="vr:Contact"/>

  </xs:sequence>
</xs:complexType>
```

---

**vr:Curation Metadata Elements**

| Element | Definition | |
|---|---|---|
| publisher | *RM Name:* | Publisher, PublisherID (see type definition below) |
| | *Value type:* | string with optional ID attribute: `vr:ResourceName` |
| | *Semantic Meaning:* | Entity (e.g. person or organisation) responsible for making the resource available |
| | *Occurrences:* | required |
| | *Comments:* | The PublisherID is encoded as an optional attribute |
| creator | *RM Name:* | Creator, Creator.Logo (see type definition below) |
| | *Value type:* | composite: `vr:Creator` |
| | *Semantic Meaning:* | The entity (e.g. person or organisation) primarily responsible for creating the content or constitution of the resource. |
| | *Occurrences:* | optional; multiple occurrences allowed |
| | *Comments:* | If `<creator>` element appears, it must contain at least a name. If multiple creator elements appear, but only one logo is provided (see type definition below), applications may assume that the one logo applies to all creators. |
| contributor | *RM Name:* | Contributor |
| | *Value type:* | string with optional ID attribute: `vr:ResourceName` |
| | *Semantic Meaning:* | Entity responsible for contributions to the content of the resource. |
| | *Occurrences:* | optional; multiple occurrences allowed |
| date | *RM Name:* | Date |
| | *Value type:* | UTC date (without time) with optional role attribute: `vr:Date` |
| | *Semantic Meaning:* | Date associated with an event in the life cycle of the resource. |
| | *Occurrences:* | optional; multiple occurrences allowed |
| version | *RM Name:* | Version |
| | *Value type:* | string: `xs:token` |
| | *Semantic Meaning:* | Label associated with creation or availablilty of a version of a resource. |
| | *Occurrences:* | optional |
| contact | *RM Name:* | Contact, Contact.Name, Contact.Email |
| | *Value type:* | composite: `vr:Contact` |

| vr:Curation Metadata Elements | |
|---|---|
| **Element** | **Definition** |
| | *Semantic Meaning:* Information that can be used for contacting someone with regard to this resource. |
| | *Occurrences:* required; multiple occurrences allowed |

Several of the curation elements (most importantly, `<publisher>`) make use of the `vr:ResourceName` type. This type is provides a means of refering to another resource both by name and by its IVOA identifier. Not all resources refered to using this type will necessarily be registered and, therefore, will have an identifier; thus, the identifier (which is encoded as an attribute) is optional.

**vr:ResourceName Type Schema Definition**

```
<xs:complexType name="ResourceName">
  <xs:simpleContent>
    <xs:extension base="xs:token">
      <xs:attribute name="ivo-id" type="vr:IdentifierURI"/>
    </xs:extension>
  </xs:simpleContent>
</xs:complexType>
```

The `<creator>` element is defined by the `vr:Creator` complex type which bundles together the RM metadata *Creator* and *Creator.Logo*.

**vr:Creator Type Schema Definition**

```
<xs:complexType name="Creator">
  <xs:sequence>
    <xs:element name="name" type="vr:ResourceName"/>
    <xs:element name="logo" type="xs:anyURI" minOccurs="0"/>
  </xs:sequence>
</xs:complexType>
```

| vr:Creator Metadata Elements | |
|---|---|
| **Element** | **Definition** |
| name | *RM Name:* Creator |
| | *Value type:* string with optional ID attribute: `vr:ResourceName` |
| | *Semantic Meaning:* The name and (optionally) the ID of the entity (e.g. person or organisation) primarily responsible for creating the content or constitution of the resource. |
| | *Occurrences:* required |
| logo | *RM Name:* Creator.Logo |
| | *Value type:* URL: `vr:paddedURI` |
| | *Semantic Meaning:* URL pointing to a graphical logo, which may be used to help identify the information source |
| | *Occurrences:* optional |

The `<Date>` element's type, `vr:Date`, is an extension of the `UTCDateTime` type that adds an optional attribute called `role`.

**vr:Date Type Schema Definition**

```
<xs:complexType name="Date">
  <xs:simpleContent>
    <xs:extension base="xs:date">
      <xs:attribute name="role" type="xs:string" default="representative"/>
    </xs:extension>
  </xs:simpleContent>
</xs:complexType>
```

The purpose of the `role` attribute is to indicate what aspect of the resource the date describes. This allows several `<date>` elements to be provided, each with a different role. The possible values for this attribute are *not* controlled; however, several values are recommended below and may be recognized by applications. Other roles may be given. If a role is not given, applications should interpret the date as `representative` as defined below.

| vr:Date Attributes | |
|---|---|
| **Attribute** | **Definition** |
| `role` | *Value type:* string: `xsd:string` |
| | *Semantic Meaning:* A string indicating what the date refers to. |
| | *Recommended Values:* `creation` date refers to when the resource was created. |
| | `update` date refers to when the resource was last updated. |
| | `representative` date is a rough representation of the time coverage of the resource (default). |

It is important to note that the `<date>` element describes the resource itself, not the resource record that describes it. Dates describing the resource record are covered by <u>vr:Resource</u> attributes `created` and `updated`.

The `<contact>` element is defined by the `vr:Contact` type which bundles together several component pieces of information, including the RM metadata *Contact.Name* and *Contact.Email*.

| vr:Contact Type Schema Definition |
|---|
| ```
<xs:complexType name="Contact">
  <xs:sequence>

    <xs:element name="name" type="vr:ResourceName"/>
    <xs:element name="address" type="xs:token" minOccurs="0"/>
    <xs:element name="email" type="xs:token" minOccurs="0"/>
    <xs:element name="telephone" type="xs:token" minOccurs="0"/>

  </xs:sequence>
</xs:complexType>
``` |

| vr:Contact Metadata Elements | |
|---|---|
| **Element** | **Definition** |
| name | *RM Name:* Contact.Name |
| | *Value type:* string with optional ID attribute: <u>vr:ResourceName</u> |
| | *Semantic Meaning:* The name and (optionally) the ID of the entity (e.g. person or organisation) that can be contacted regarding the resource. |
| | *Occurrences:* required |
| address | *RM Name:* *n. a.* |
| | *Value type:* string: `xs:token` |
| | *Semantic Meaning:* All components of the mailing address of the contact person or organisation. |
| | *Occurrences:* optional |
| email | *RM Name:* Contact.Email |
| | *Value type:* string: `xs:token` |
| | *Semantic Meaning:* The email address of the contact person or organisation. |
| | *Occurrences:* optional |
| telephone | *RM Name:* *n. a.* |
| | *Value type:* string: `xs:token` |
| | *Semantic Meaning:* The telephone number of the contact person or organisation. |
| | *Occurrences:* optional |
| | *Comments:* Complete international dialing codes should be provided. |

### 3.1.3. General Content Metadata



The general content metadata described in the RM (section 3.3) are bundled together into the `vr:Resource` child element, `<content>`. Its content is defined by the `vr:Content` complex type.

---

**vr:Content Type Schema Definition**

```
<xs:complexType name="Content">
  <xs:sequence>

    <xs:element name="subject" type="xs:token"
      minOccurs="0" maxOccurs="unbounded"/>
    <xs:element name="description" type="xs:token"/>
    <xs:element name="source" type="vr:Source" minOccurs="0"/>
    <xs:element name="referenceURL" type="vr:PaddedURI"/>
    <xs:element name="type" type="vr:Type" minOccurs="0"/>
    <xs:element name="contentLevel" type="vr:ContentLevel"
      minOccurs="0" maxOccurs="unbounded"/>
    <xs:element name="relationship" type="vr:Relationship"
      minOccurs="0" maxOccurs="unbounded"/>

  </xs:sequence>
</xs:complexType>
```

---

**vr:Content Metadata Elements**

| Element | Definition | |
|---|---|---|
| subject | *RM Name:* | Subject |
| | *Value type:* | string: `xs:token` |
| | *Semantic Meaning:* | a topic, object type, or other descriptive keywords about the resource. |
| | *Occurrences:* | required; multiple occurrences allowed |
| description | *RM Name:* | Description |
| | *Value type:* | string: `xs:token` |
| | *Semantic Meaning:* | An account of the nature of the resource. |
| | *Occurrences:* | required |
| | *Comments:* | The description may include but is not limited to an abstract, table of contents, reference to a graphical representation of content or a free-text account of the content. |
| source | *RM Name:* | Source |
| | *Value type:* | a formatted reference (e.g. ADS Bibcode): `vr:Source` |
| | *Semantic Meaning:* | a bibliographic reference from which the present resource is derived or extracted. |
| | *Occurrences:* | optional |
| referenceURL | *RM Name:* | ReferenceURL |
| | *Value type:* | URL: `vr:PaddedURI` |
| | *Semantic Meaning:* | URL pointing to a human-readable document describing the resource. |
| | *Occurrences:* | required |
| type | *RM Name:* | Type |
| | *Value type:* | string with controlled vocabulary: `vr:Type` |
| | *Semantic Meaning:* | Nature or genre of the content of the resource |
| | *Occurrences:* | optional; multiple occurrences allowed |
| | *Allowed Values:* | `Other`  resource that does not fall into any of the category names currently defined. |
| | | `Archive`  Collection of pointed observations |
| | | `Bibliography`  Collection of bibliographic reference, abstracts, and publications |
| | | `Catalog`  Collection of derived data, primarily in tabular form |
| | | `Journal`  Collection of scholarly publications under common editorial policy |



**vr:Content Metadata Elements**

| Element | Definition | |
|---|---|---|
| | `Library` | Collection of published materials (journals, books, etc.) |
| | `Simulation` | Theoretical simulation or model |
| | `Survey` | Collection of observations covering substantial and contiguous areas of the sky |
| | `Transformation` | A service that transforms data |
| | `Education` | Collection of materials appropriate for educational use, such as teaching resources, curricula, etc. |
| | `Outreach` | Collection of materials appropriate for public outreach, such as press releases and photo galleries |
| | `EPOResource` | Collection of materials that may be suitable for EPO products but which are not in final product form, as in Type Outreach or Type Education. EPOResource would apply, e.g., to archives with easily accessed preview images or to surveys with easy-to-use images. |
| | `Animation` | Animation clips of astronomical phenomena |
| | `Artwork` | Artists' renderings of astronomical phenomena or objects |
| | `Background` | Background information on astronomical phenomena or objects |
| | `BasicData` | Compilations of basic astronomical facts about objects, such as approximate distance or membership in constellation |
| | `Historical` | Historical information about astronomical objects |
| | `Photographic` | Publication-quality photographs of astronomical objects |
| | `Press` | Press releases about astronomical objects |
| | `Organisation` | An organization that is a publisher or curator of other resources |
| | `Project` | A project that is a publisher or curator of other resources |
| | `Registry` | a query service for which the response is a structured description of resources |
| contentLevel | *RM Name:* | ContentLevel |
| | *Value type:* | string with controlled vocabulary: `vr:ContentLevel` |
| | *Semantic Meaning:* | Description of the content level or intended audience |
| | *Occurrences:* | optional; multiple occurrences allowed |
| | *Allowed Values:* | `General`   Resource provides information appropriate for all users. |
| | `Elementary Education` | Resource provides information appropriate for use in elementary education (e.g. approximate ages 6-11). |
| | `Middle School Education` | Resource provides information appropriate for use in middle school education (e.g. approximate ages 11-14). |
| | `Secondary Education` | Resource provides information appropriate for use in elementary education (e.g. approximate ages 14-18). |



| vr:Content Metadata Elements | | |
|---|---|---|
| **Element** | **Definition** | |

| | `Community College` | Resource provides information appropriate for use in community/junior college or early university education. |
|---|---|---|
| | `University` | Resource provides information appropriate for use in university education. |
| | `Research` | Resource provides information appropriate for supporting scientific research. |
| | `Amateur` | Resource provides information of interest to amateur astronomers. |
| | `Informal Education` | Resource provides information appropriate for education at museums, planetariums, and other centers of informal learning. |
| relationship | *RM Name:* | Relationship, RelationshipID |
| | *Value type:* | composite: `vr:Relationship` |
| | *Semantic Meaning:* | a description of a relationship to another resource |
| | *Occurrences:* | optional; multiple occurrences allowed |

The `<source>` element's type, `vr:Source`, is an extension of the `xs:token` type that adds an optional attribute called `format`.

| vr:Source Type Schema Definition |
|---|

```
<xs:complexType name="Source">
    <xs:simpleContent>
        <xs:extension base="xs:token">
            <xs:attribute name="format" type="xs:string"/>
        </xs:extension>
    </xs:simpleContent>
</xs:complexType>
```

The `format` indicates the syntactic format of the value of the `<source>` element. The possible values for this attribute are *not* controlled; however, applications should recognize a value equal to `bibcode` as referring to the standard astronomical bibcode format [Bibcode].

The `<relationship>` is defined by the `vr:Relationship` complex type which bundles together the RM metadata *Relationship* and *RelationshipID*.

| vr:Relationship Type Schema Definition |
|---|

```
<xs:complexType name="Relationship">
    <xs:sequence>
        <xs:element name="relationshipType" type="xs:token"/>
        <xs:element name="relatedResource" type="vr:ResourceName"
                    minOccurs="1" maxOccurs="unbounded"/>
    </xs:sequence>
</xs:complexType>
```

| vr:Relationship Metadata Elements | | |
|---|---|---|
| **Element** | **Definition** | |
| relatedResource | *RM Name:* | RelationshipID |
| | *Value type:* | string with optional ID attribute: `vr:ResourceName` |
| | *Semantic Meaning:* | the name of resource that this resource is related to. |
| | *Occurrences:* | required, multiple values allowed |
| | *Comments:* | The attribute value provides the RM metadatum *RelationshipID*. |
| relationshipType | *RM Name:* | Relationship |
| | *Value type:* | string with controlled vocabulary |
| | *Semantic Meaning:* | the named type of relationship |



**vr:Relationship Metadata Elements**

| Element | Definition | |
|---|---|---|
| | *Occurrences:* | required |
| | *Allowed Values:* | `mirror-of` The current resource mirrors another resource. |
| | | `service-for` The current resource is a service that provides access to a data collection. |
| | | `served-by` The current resource can be accessed via the identified service. |
| | | `derived-from` The current resource is derived from another resource. |
| | | `related-to` The current resource is related to another resource in an unspecified way. |
| | *Comments:* | The allowed values are not enforced by the schema itself. The "other" resource refered to in the above definitions refers to the resource given by the `<relatedResource>` element. |

It is important to note that the `<relationshipType>` controlled vocabulary is not enforced by the schema itself. This allows a future version of this specification to define additional terms without requiring a change in the XML Schema document.

### 3.1.4. Resource Record Quality with validationLevel

The RM's *ResourceValidationLevel* is encoded in the VOResource schema with the `<validationLevel>` element, which can appear in multiple places within a `vr:Resource` type or sub-type. It may appear one or more times as the first children of a `vr:Resource` type and/or, if the resource is a `vr:Service` type or sub-type, one or more times as the first children of any `<capability>` element.

**The `validationLevel` Element**

| Element | Definition | |
|---|---|---|
| validationLevel | *RM Name:* | ResourceValidationLevel |
| | *Value type:* | integer with required `validatedBy` attribute taking an ID: `vr:Validation` |
| | *Semantic Meaning:* | An evaluation via a numeric grade that indicates the quality of the resource description, when applicable, to be used to indicate the confidence an end-user can put in the resource as part of a VO application or research study. |
| | *Occurrences:* | optional; multiple occurrences allowed |
| | *Comments:* | The part of the resource record that this evaluation applies to depends on whether this element appears as a child of a `<capability>` element or as a direct child of a `vr:Resource` record; see discussion below. |
| | *Allowed Values:* | `0` The resource has a description that is stored in a registry. This level does not imply a compliant description. |
| | | `1` In addition to meeting the level 0 definition, the resource description conforms syntactically to this standard and to the encoding scheme used. |
| | | `2` In addition to meeting the level 1 definition, the resource description refers to an existing resource that has demonstrated to be functionally compliant. |
| | | When the resource is a service, it is consider to exist and functionally compliant if use of the service accessURL responds without error when used as intended by the resource. If the service is a standard one, it must also demonstrate the response is syntactically compliant with the service standard in order to be considered functionally |



| The `validationLevel` Element | |
|---|---|
| **Element** | **Definition** |
| | compliant. If the resource is not a service, then the ReferenceURL must be shown to return a document without error. |
| | 3  In addition to meeting the level 2 definition, the resource description has been inspected by a human and judged to comply semantically to this standard as well as meeting any additional minimum quality criteria (e.g., providing values for important but non-required metadata) set by the human inspector. |
| | 4  In addition to meeting the level 3 definition, the resource description meets additional quality criteria set by the human inspector and is therefore considered an excellent description of the resource. Consequently, the resource is expected to be operate well as part of a VO application or research study. |

The `validationLevel` element requires an attribute called `validatedBy` which takes an IVOA ID as a value. This ID must refer to a registered organisation or registry that assigned the numerical value. This element can appear multiple times, each with a different `validatedBy` value, to reflect the code assigned by different organisations.

**vr:Validation Type Schema Definition**

```
<xs:complexType name="Validation">
  <xs:simpleContent>
    <xs:extension base="vr:ValidationLevel">
      <xs:attribute name="validatedBy" type="vr:IdentifierURI"
               use="required"/>
    </xs:extension>
  </xs:simpleContent>
</xs:complexType>

<xs:simpleType name="ValidationLevel">
  <xs:restriction base="xs:integer">
    <xs:whiteSpace value="collapse"/>
    <xs:enumeration value="0"/>
    <xs:enumeration value="1"/>
    <xs:enumeration value="2"/>
    <xs:enumeration value="3"/>
    <xs:enumeration value="4"/>
  </xs:restriction>
</xs:simpleType>
```

## 3.2. Resource Type Extensions: Organisation and Service

The VOResource schema defines two extensions of the base `vr:Resource` type for describing two specific types of resources: `vr:Organisation` and `vr:Service`. In addition to providing more refined semantic meaning over `vr:Resource`, they add additional metadata for the describing the resource which don't necessarily apply in the generic case.

### 3.2.1. The Organisation Resource Type

The Organisation resource type is used to describe an organisation in the sense defined by the RM:

> An organisation is a specific type of resource that brings people together to pursue participation in VO applications. Organisations can be hierarchical and range greatly in size and scope. At a high level, an organisation could be a university, observatory, or government agency. At a finer level, it could be a specific scientific project space mission, or individual researcher.

The Organisation type extends the Resource type by adding two additional elements to the core set of metadata, **`<facility>`** and **`<instrument>`**:

**vr:Organisation Type Schema Definition**

```
<xs:complexType name="Organisation">
```



```
        <xs:complexContent>
            <xs:extension base="vr:Resource">
                <xs:sequence>

                    <xs:element name="facility" type="vr:ResourceName"
                                minOccurs="0" maxOccurs="unbounded"/>
                    <xs:element name="instrument" type="vr:ResourceName"
                                minOccurs="0" maxOccurs="unbounded"/>

                </xs:sequence>
            </xs:extension>
        </xs:complexContent>
    </xs:complexType>
```

As outlined below, both elements take a name as a value; however, an IVOA Identifier may also be provided as an attribute if the facility or instrument mentioned has a registered description associated with it.

| vr:Organisation Extension Metadata Elements | | |
|---|---|---|
| Element | Definition | |
| facility | *RM Name:* | Facility |
| | *Value type:* | string with optional ID attribute: `vr:ResourceName` |
| | *Semantic Meaning:* | the observatory or facility used to collect the data contained or managed by this resource. |
| | *Occurrences:* | optional; multiple occurrences allowed |
| instrument | *RM Name:* | Instrument |
| | *Value type:* | string with optional ID attribute: `vr:ResourceName` |
| | *Semantic Meaning:* | the instrument used to collect the data contained or managed by this resource. |
| | *Occurrences:* | optional; multiple occurrences allowed |

The main role of an organisation in the VO (that is, the main reason for describing organisations in a registery) is as a provider or publisher of other resources. In particular, an organisation description in a registry declares the association of an IVOA identifier [ID] with the organisation. The organisation can then be referred to in other resource descriptions. For example, an organisation identifier will appear as the publisher identifier of service resource (as illustrated in our example from section 2).

### 3.2.2. The Service Resource Type

The Service resource type is used to describe a service--a resource that actually does something--in the sense defined by the RM:

> A service is any VO resource that can be invoked by the user to perform some action on their behalf.

The general data model is described in section 2.2.2. The Service type extends the Resource type by adding two elements: `<rights>` which indicates who may access it, and `<capability>` which describes how the service behaves and how it is invoked.

| vr:Service Type Schema Definition |
|---|
| ```
<xs:complexType name="Service">
    <xs:complexContent>
        <xs:extension base="vr:Resource">
            <xs:sequence>
                <xs:element name="rights" type="vr:Rights"
                            minOccurs="0" maxOccurs="unbounded"/>
                <xs:element name="capability" type="vr:Capability"
                            minOccurs="0" maxOccurs="unbounded"/>
            </xs:sequence>
        </xs:extension>
    </xs:complexContent>
</xs:complexType>
``` |

| vr:Service Extension Metadata Elements | |
|---|---|
| Element | Definition |

**vr:Service Extension Metadata Elements**

| Element | Definition |
|---|---|
| rights | *RM Name:* Rights |
| | *Value type:* string, controlled vocabulary: `xs:token` |
| | *Semantic Meaning:* Information about rights held in and over the resource. |
| | *Occurrences:* optional; multiple occurrences allowed |
| | *Allowed Values:* `public` unrestricted, public access is allowed without authentication. |
| | `secure` authenticated, public access is allowed. |
| | `proprietary` only proprietary access is allowed with authentication. |
| capability | *Value type:* composite: `vr:Capability` |
| | *Semantic Meaning:* a description of a general capability of the service--its behavioral characteristics and limitations--and how to use it. |
| | *Occurrences:* optional; multiple occurrences allowed |

As described in section 2.2.2, multiple `capability` elements may appear, each describing a different capability of the service.

**vr:Capability Type Schema Definition**

```
<xs:complexType name="Capability">
    <xs:sequence>
        <xs:element name="validationLevel" type="vr:Validation"
            minOccurs="0" maxOccurs="unbounded"/>
        <xs:element name="description" type="xs:token" minOccurs="0"/>
        <xs:element name="interface" type="vr:Interface"
            minOccurs="0" maxOccurs="unbounded"/>
    </xs:sequence>
    <xs:attribute name="standardID" type="xs:anyURI"/>
</xs:complexType>
```

**vr:Capability Metadata Elements**

| Element | Definition |
|---|---|
| validationLevel | *RM Name:* ResourceValidationLevel |
| | *Value type:* integer with required `validatedBy` attribute taking an ID: `vr:Validation` (see definition for listing of allowed values.) |
| | *Semantic Meaning:* An evaluation via a numeric grade that indicates the quality of the capability description and whether its implementation is functionally with applicable standards. |
| | *Occurrences:* optional; multiple occurrences allowed |
| description | *Value type:* string: `xs:token` |
| | *Semantic Meaning:* A human-readable description of what this capability provides as part of the over-all service |
| | *Occurrences:* optional |
| interface | *Value type:* composite: `vr:Interface` |
| | *Semantic Meaning:* a description of how to call the service to access this capability |
| | *Occurrences:* optional |

The `capability` element is sufficient for describing a *custom service capability*--i.e., a service that is particular to one provider and does not conform to a specific standard (be it recognized by the IVOA or some other sub-community). However, service standards will often create a `vr:Capability` extension that adds additional metadata that are specific that that are specific to the behavior of that particular type of service.

> **Note:**
> The RM defines three metadata that may be important for several standard query

> services: *Service.MaxSearchRadius*, *Service.MaxReturnRecords*, and *Service.MaxReturnSize*. These are examples of service-specific metadata that should be encoded as child elements in a type derived from `vr:Capability`.

| vr:Capability Attribute | |
|---|---|
| **Attribute** | **Definition** |
| `standardID` | *RM Name:* Service.StandardURI<br>*Value type:* a URI: `xs:anyURI`<br>*Semantic Meaning:* A unique identifier for the standard that the service complies to.<br>*Comments:* Normally this attribute is optional; however, a `vr:Capability` extension may force this attribute to be required and set to a fixed URI representing the service standard. |

The `<interface>` element is of the complex type `vr:Interface`.

| vr:Interface Type Schema Definition |
|---|
| ```<br><xs:complexType name="Interface" abstract="true"><br>    <xs:sequence><br>        <xs:element name="accessURL" type="vr:AccessURL"<br>            maxOccurs="unbounded"/><br>        <xs:element name="securityMethod" type="vr:SecurityMethod"<br>            minOccurs="0" maxOccurs="unbounded"/><br>    </xs:sequence><br><br>    <xs:attribute name="version" type="xs:string" default="1.0"/><br>    <xs:attribute name="role" type="xs:NMTOKEN"/><br></xs:complexType><br>``` |

The `vr:Interface` type is defined as abstract, so as described in , the `<interface>` element must not be used as part of a `vr:Service` description unless it includes an `xsi:type` attribute that refers to a type derived from `vr:Interface`. The VOResource schema defines two derived `vr:Interface` types: vr:WebBrowser and vr:WebService.

As described in , the `vr:Interface` attributes help distinguish between multiple interfaces within the same `<capability>` element:

| vr:Interface Attributes | |
|---|---|
| **Attribute** | **Definition** |
| `version` | *Value type:* string: `xs:string`<br>*Semantic Meaning:* The version of a standard interface specification that this interface complies with. When the interface is provided in the context of a Capability element, then the standard being refered to is the one identified by the Capability's standardID element. If the standardID is not provided, the meaning of this attribute is undefined.<br>*Default Value:* 1.0 |
| `role` | *Value type:* string: `xs:NMTOKEN`<br>*Semantic Meaning:* A tag name the identifies the role the interface plays in the particular capability. If the value is equal to "std" or begins with "std:", then the interface refers to a standard interface defined by the standard referred to by the capability's standardID attribute.<br>*Comments:* Prefer "std" if there is only one interface mandated by the standard. |

The `vr:Interface` type provides two child elements: `<accessURL>` and `<securityMethod>`.

| vr:Interface Metadata Elements | |
|---|---|
| **Element** | **Definition** |

| vr:Interface Metadata Elements | |
|---|---|
| **Element** | **Definition** |
| accessURL | *RM Name:* ServiceURL, BaseURL |
| | *Value type:* URL with <u>optional use attribute</u>: `vr:AccessURL` |
| | *Semantic Meaning:* The URL (or base URL) that a client uses to access the service. How this URL is to be interpreted and used depends on the specific Interface subclass. |
| | *Occurrences:* required; multiple occurrences allowed. |
| securityMethod | *Value type:* composite: <u>a sub-type of</u> `vr:SecurityMethod` |
| | *Semantic Meaning:* the mechanism the client must employ to gain secure access to the service. |
| | *Occurrences:* optional; multiple occurrences allowed. |

As mentioned in the table above, exactly how a client uses the value of the `<accessURL>` element depends on the specific type derived from `vr:Interface`. Extension schemas that define non-abstract types derived from `vr:Interface` MUST provide documentation that explains the exact use of the `<accessURL>`; this documentation should follow the <u>documention conventions</u> described in <u>section 2.2</u>. If multiple `<accessURL>` may be provide; however, each URL should point to a functionally identical or "mirror" installation of the same service and administered by the same publisher listed above. An `<accessURL>` must not point to an installation that is outside the administrative control of the service's listed publisher; such a mirror should be describe in a separate resource record.

As an additional aid to software agents that will attempt to interpret its URL, the `vr:AccessURL` includes an optional element, `use` (pronounced and interpreted as in the noun form of the word use) which takes a controlled vocabulary:

| vr:AccessURL Attribute | |
|---|---|
| **Attribute** | **Definition** |
| `use` | *Value type:* string, controlled vocabulary: `xsd:string` |
| | *Semantic Meaning:* A flag indicating the general way the URL is used. |
| | *Allowed Values:* |
| |    `full` Assume a full URL--that is, one that can be invoked directly without alteration. This usually returns a single document or file. |
| |    `base` Assume a base URL--that is, one requiring an extra portion to be appended before being invoked. This value should be provide for services that used the HTTP GET mechanism. |
| |    `post` Assume the URL is a service endpoint that requires input sent via the HTTP POST mechanism. This is the value that should be provide for HTTP SOAP web services. |
| |    `dir` Assume URL points to a directory that will return a listing of files. |

| vr:AccessURL Type Schema Definition |
|---|
| ```xml
<xs:complexType name="AccessURL">
  <xs:simpleContent>
    <xs:extension base="xs:anyURI">
      <xs:attribute name="use">
        <xs:simpleType>
          <xs:restriction base="xs:string">
            <xs:enumeration value="full"/>
            <xs:enumeration value="base"/>
            <xs:enumeration value="post"/>
            <xs:enumeration value="dir"/>
          </xs:restriction>
        </xs:simpleType>
      </xs:attribute>
    </xs:extension>
  </xs:simpleContent>
</xs:complexType>
``` |

The `vr:SecurityMethod` type is defined as a complex type but with empty content:



| vr:SecurityMethod Type Schema Definition |
|---|

```
<xs:complexType name="SecurityMethod">
    <xs:sequence/>
    <xs:attribute name="standardID" type="xs:anyURI"/>
</xs:complexType>
```

| vr:Capability Attribute | |
|---|---|
| **Attribute** | **Definition** |
| `standardID` | *Value type:*      a URI: `xs:anyURI`<br>*Semantic Meaning:* A unique identifier for a security mechanism supported by this interface. |

While this simple element (when the `standardID` attribute is provided) may on its own be sufficient to describe the security mechanism used, it is expected that some future VOResource extension schema will define additional types derived from `vr:SecurityMethod`. If such a sub-type is available, it may be employed at `securityMethod` location within a `vr:Interface`-typed element, in which case, it should be invoked via a `xsi:type` attribute to the `securityMethod` element.

`vr:WebBrowser` is one of the two `vr:Interface` sub-types defined by the VOResource schema. This type indicates that the service is intended to be accessed interactively by a user through a web browser. The `<accessURL>`, then, reprensents the URL of a web page containing one or more forms used to invoke the service.

| vr:WebBrowser Type Schema Definition |
|---|

```
<xs:complexType name="WebBrowser">
    <xs:complexContent>
        <xs:extension base="vr:Interface">
            <xs:sequence/>
        </xs:extension>
    </xs:complexContent>
</xs:complexType>
```

As can be seen in the schema definition above, the `vr:WebBrowser` type does not define any additional interface metadata (though other `vr:Interface` derivations may). Thus, this type is provide purely for its semantic meaning.

`vr:WebService` is the second `vr:Interface` sub-type available from the VOResource schema:

| vr:WebService Type Schema Definition |
|---|

```
<xs:complexType name="WebService">
    <xs:complexContent>
        <xs:extension base="vr:Interface">
            <xs:sequence>
                <xs:element name="wsdlURL" type="xs:anyURI"
                            minOccurs="0" maxOccurs="unbounded"/>
            </xs:sequence>
        </xs:extension>
    </xs:complexContent>
</xs:complexType>
```

The `vr:WebService` interface is one that is described by a Web Service Description Language [WSDL] document. This is typically realized as one that employs the Simple Object Access Protocol [SOAP]; however, like WSDL, this interface type is not restricted to it. With this interface, the `vr:accessURL` must refer to the service endpoint URL.

| vr:WebService Extension Metadata Elements | |
|---|---|
| **Element** | **Definition** |
| wsdlURL | *Value type:*      a URL: `xs:anyURI`<br>*Semantic Meaning:* The location of the WSDL that describes this Web Service. If not provided, the location is assumed to be the access URL with "?wsdl" appended.<br>*Occurrences:*      optional; multiple occurrences allowed |



| vr:WebService Extension Metadata Elements | | |
|---|---|---|
| **Element** | | **Definition** |
| | *Comments:* | Multiple occurrences should represent mirror copies of the same WSDL file. |

> **Note:**
> It is intended that other `vr:Interface` types, along with additional resource types derived from `vr:Service`, will be defined in at least one other IVOA standard extension that is specifically geared to services.

## Appendix A: the VOResource XML Schema

| The Complete VOResource Schema |
|---|

```
<?xml version="1.0" encoding="UTF-8"?>
<xs:schema targetNamespace="http://www.ivoa.net/xml/VOResource/v1.0"
        xmlns="http://www.w3.org/2001/XMLSchema"
        xmlns:xs="http://www.w3.org/2001/XMLSchema"
        xmlns:vr="http://www.ivoa.net/xml/VOResource/v1.0"
        xmlns:vm="http://www.ivoa.net/xml/VOMetadata/v0.1"
        elementFormDefault="unqualified"
        attributeFormDefault="unqualified"
        version="1.0r7">

    <xs:annotation>
      <xs:appinfo>
        <vm:schemaName>VOResource</vm:schemaName>
        <vm:schemaPrefix>xs</vm:schemaPrefix>
        <vm:targetPrefix>vr</vm:targetPrefix>
      </xs:appinfo>
      <xs:documentation>
        Implementation of an XML Schema describing a resource to
        be used in the Virtual Observatory Project.  Based on "Resource
        Metadata for the Virtual Observatory", Version 0.8,
        February 2002 by Bob Hanisch et al.
      </xs:documentation>
    </xs:annotation>

    <xs:simpleType name="UTCTimestamp">
        <xs:annotation>
            <xs:documentation>
            A timestamp that is compliant with ISO8601 but disallows
            the use of a timezone indicator.
            </xs:documentation>
        </xs:annotation>

        <xs:restriction base="xs:dateTime">
            <xs:pattern value="\d{4}-\d\d-\d\dT\d\d:\d\d:\d\d(\.\d+)?"/>
        </xs:restriction>
    </xs:simpleType>

    <xs:simpleType name="UTCDateTime">
        <xs:annotation>
            <xs:documentation>
            A date stamp that can be given to a precision of either a
            day (type xs:date) or seconds (type
            xs:dateTime)
            </xs:documentation>
        </xs:annotation>
        <xs:union memberTypes="xs:date vr:UTCTimestamp"/>
    </xs:simpleType>

    <xs:complexType name="Resource">
        <xs:annotation>
            <xs:documentation>
            Any entity or component of a VO application that is
            describable and identifiable by a IVOA Identifier.
            </xs:documentation>
        </xs:annotation>
        <xs:sequence>
            <xs:element name="validationLevel" type="vr:Validation"
                    minOccurs="0" maxOccurs="unbounded">
                <xs:annotation>
                    <xs:documentation>
                    A numeric grade describing the quality of the
                    resource description, when applicable,
                    to be used to indicate the confidence an end-user
                    can put in the resource as part of a VO application
                    or research study.
                    </xs:documentation>
                    <xs:documentation>
                    See vr:ValidationLevel for an explanation of the
                    allowed levels.
                    </xs:documentation>
                    <xs:documentation>
                    Note that when this resource is a Service, this
                    grade applies to the core set of metadata.
```



```
                    Capability and interface metadata, as well as the
                    compliance of the service with the interface
                    standard, is rated by validationLevel tag in the
                    capability element (see the vr:Service complex
                    type).
                </xs:documentation>
            </xs:annotation>
        </xs:element>

        <xs:element name="title" type="xs:token">
            <xs:annotation>
                <xs:appinfo>
                    <vm:dcterm>Title</vm:dcterm>
                </xs:appinfo>
                <xs:documentation>
                    the full name given to the resource
                </xs:documentation>
            </xs:annotation>
        </xs:element>

        <xs:element name="shortName" type="vr:ShortName" minOccurs="0">
            <xs:annotation>
                <xs:documentation>
                    a short name or abbreviation given to the resource.
                </xs:documentation>
                <xs:documentation>
                    This name will be used where brief annotations for
                    the resource name are required.  Applications may
                    use to refer to this resource in a compact display.
                </xs:documentation>
                <xs:documentation>
                    One word or a few letters is recommended.  No more
                    than sixteen characters are allowed.
                </xs:documentation>
            </xs:annotation>
        </xs:element>

        <xs:element name="identifier" type="vr:IdentifierURI">
            <xs:annotation>
                <xs:appinfo>
                    <vm:dcterm>Identifier</vm:dcterm>
                </xs:appinfo>
                <xs:documentation>
                    Unambiguous reference to the resource conforming to the IVOA
                    standard for identifiers
                </xs:documentation>
            </xs:annotation>
        </xs:element>

        <xs:element name="curation" type="vr:Curation">
            <xs:annotation>
                <xs:documentation>
                    Information regarding the general curation of the resource
                </xs:documentation>
            </xs:annotation>
        </xs:element>

        <xs:element name="content" type="vr:Content">
            <xs:annotation>
                <xs:documentation>
                    Information regarding the general content of the resource
                </xs:documentation>
            </xs:annotation>
        </xs:element>

    </xs:sequence>

    <xs:attribute name="created" type="xs:dateTime" use="required">
        <xs:annotation>
            <xs:documentation>
                The UTC date and time this resource metadata description
                was created.
            </xs:documentation>
            <xs:documentation>
                This timestamp must not be in the future.  This time is
                not required to be accurate; it should be at least
                accurate to the day.  Any insignificant time fields
                should be set to zero.
            </xs:documentation>
        </xs:annotation>
    </xs:attribute>

    <xs:attribute name="updated" type="xs:dateTime" use="required">
        <xs:annotation>
            <xs:documentation>
                The UTC date this resource metadata description was last updated.
            </xs:documentation>
            <xs:documentation>
                This timestamp must not be in the future.  This time is
                not required to be accurate; it should be at least
                accurate to the day.  Any insignificant time fields
                should be set to zero.
            </xs:documentation>
        </xs:annotation>
    </xs:attribute>

    <xs:attribute name="status" default="active">
        <xs:annotation>
            <xs:documentation>
                a tag indicating whether this resource is believed to be still
                actively maintained.
            </xs:documentation>
        </xs:annotation>
```



```xml
                <xs:simpleType>
                    <xs:restriction base="xs:string">
                        <xs:enumeration value="active">
                            <xs:annotation>
                                <xs:documentation>
                                    resource is believed to be currently maintained, and its
                                    description is up to date (default).
                                </xs:documentation>
                            </xs:annotation>
                        </xs:enumeration>
                        <xs:enumeration value="inactive">
                            <xs:annotation>
                                <xs:documentation>
                                    resource is apparently not being maintained at the present.
                                </xs:documentation>
                            </xs:annotation>
                        </xs:enumeration>
                        <xs:enumeration value="deleted">
                            <xs:annotation>
                                <xs:documentation>
                                    resource publisher has explicitly deleted the resource.
                                </xs:documentation>
                            </xs:annotation>
                        </xs:enumeration>
                    </xs:restriction>
                </xs:simpleType>
            </xs:attribute>
    </xs:complexType>

    <xs:simpleType name="ValidationLevel">
        <xs:annotation>
            <xs:documentation>
                the allowed values for describing the resource descriptions
                and interfaces.
            </xs:documentation>
            <xs:documentation>
                See the RM (v1.1, section 4) for more guidance on the use of
                these values.
            </xs:documentation>
        </xs:annotation>
        <xs:restriction base="xs:integer">
            <xs:whiteSpace value="collapse"/>
            <xs:enumeration value="0">
                <xs:annotation>
                    <xs:documentation>
                        The resource has a description that is stored in a
                        registry. This level does not imply a compliant
                        description.
                    </xs:documentation>
                </xs:annotation>
            </xs:enumeration>
            <xs:enumeration value="1">
                <xs:annotation>
                    <xs:documentation>
                        In addition to meeting the level 0 definition, the
                        resource description conforms syntactically to this
                        standard and to the encoding scheme used.
                    </xs:documentation>
                </xs:annotation>
            </xs:enumeration>
            <xs:enumeration value="2">
                <xs:annotation>
                    <xs:documentation>
                        In addition to meeting the level 1 definition, the
                        resource description refers to an existing resource that
                        has demonstrated to be functionally compliant.
                    </xs:documentation>
                    <xs:documentation>
                        When the resource is a service, it is consider to exist
                        and functionally compliant if use of the
                        service accessURL responds without error when used as
                        intended by the resource. If the service is a standard
                        one, it must also demonstrate the response is syntactically
                        compliant with the service standard in order to be
                        considered functionally compliant. If the resource is
                        not a service, then the ReferenceURL must be shown to
                        return a document without error.
                    </xs:documentation>
                </xs:annotation>
            </xs:enumeration>
            <xs:enumeration value="3">
                <xs:annotation>
                    <xs:documentation>
                        In addition to meeting the level 2 definition, the
                        resource description has been inspected by a human and
                        judged to comply semantically to this standard as well
                        as meeting any additional minimum quality criteria (e.g.,
                        providing values for important but non-required
                        metadata) set by the human inspector.
                    </xs:documentation>
                </xs:annotation>
            </xs:enumeration>
            <xs:enumeration value="4">
                <xs:annotation>
                    <xs:documentation>
                        In addition to meeting the level 3 definition, the
                        resource description meets additional quality criteria
                        set by the human inspector and is therefore considered
                        an excellent description of the resource. Consequently,
                        the resource is expected to be operate well as part of a
                        VO application or research study.
                    </xs:documentation>
                </xs:annotation>
```



```
          </xs:enumeration>
        </xs:restriction>
  </xs:simpleType>

  <xs:complexType name="Validation">
    <xs:annotation>
      <xs:documentation>
        a validation stamp combining a validation level and the ID of
        the validator.
      </xs:documentation>
    </xs:annotation>
    <xs:simpleContent>
      <xs:extension base="vr:ValidationLevel">
        <xs:attribute name="validatedBy" type="vr:IdentifierURI"
                      use="required">
         <xs:annotation>
            <xs:documentation>
               The IVOA ID of the registry or organisation that
               assigned the validation level.
            </xs:documentation>
          </xs:annotation>
        </xs:attribute>
      </xs:extension>
    </xs:simpleContent>
  </xs:complexType>

  <xs:simpleType name="AuthorityID">
    <xs:restriction base="xs:string">
      <xs:pattern value="[\w\d][\w\d\-_\.!~\*'\(\)\+=]{2,}"/>
    </xs:restriction>
  </xs:simpleType>

  <xs:simpleType name="ResourceKey">
    <xs:restriction base="xs:string">
      <xs:whiteSpace value="collapse"/>
      <xs:pattern value="[\w\d\-_\.!~\*'\(\)\+=]+(/[\w\d\-_\.!~\*'\(\)\+=]+)*"/>
    </xs:restriction>
  </xs:simpleType>

  <xs:simpleType name="IdentifierURI">
    <xs:restriction base="xs:anyURI">
      <xs:whiteSpace value="collapse"/>
      <xs:pattern value="ivo://[\w\d][\w\d\-_\.!~\*'\(\)\+=]{2,}(/[\w\d\-_\.!~\*'\(\)\+=]+(/[\w\d\-_\.!~\*'\(\)\+=]+)*)?"/>
    </xs:restriction>
  </xs:simpleType>

  <xs:simpleType name="ShortName">
    <xs:annotation>
      <xs:documentation>
        a short name or abbreviation given to something.
      </xs:documentation>
      <xs:documentation>
        This name will be used where brief annotations for
        the resource name are required.  Applications may
        use to refer to this resource in a compact display.
      </xs:documentation>
      <xs:documentation>
        One word or a few letters is recommended.  No more
        than sixteen characters are allowed.
      </xs:documentation>
    </xs:annotation>

    <xs:restriction base="xs:string">
       <xs:whiteSpace value="collapse"/>
      <xs:maxLength value="16"/>
    </xs:restriction>
  </xs:simpleType>

  <xs:complexType name="Curation">
    <xs:annotation>
      <xs:documentation>
        Information regarding the general curation of a resource
      </xs:documentation>
    </xs:annotation>

    <xs:sequence>
      <xs:element name="publisher" type="vr:ResourceName">
        <xs:annotation>
          <xs:appinfo>
            <vm:dcterm>Publisher</vm:dcterm>
          </xs:appinfo>
          <xs:documentation>
            Entity (e.g. person or organisation) responsible for making the
            resource available
          </xs:documentation>
        </xs:annotation>
      </xs:element>

      <xs:element name="creator" type="vr:Creator"
                  minOccurs="0" maxOccurs="unbounded">
        <xs:annotation>
          <xs:appinfo>
            <vm:dcterm>Creator</vm:dcterm>
          </xs:appinfo>
          <xs:documentation>
            The entity (e.g. person or organisation) primarily responsible
            for creating the content or constitution of the resource.
          </xs:documentation>
          <xs:documentation>
            A logo need only be provided for the first occurance.
            When multiple logos are supplied via multiple creator
            elements, the application is free to choose which to
            use.
```



```
                        </xs:documentation>
                    </xs:annotation>
                </xs:element>

                <xs:element name="contributor" type="vr:ResourceName"
                            minOccurs="0" maxOccurs="unbounded">
                    <xs:annotation>
                        <xs:appinfo>
                            <vm:dcterm>Contributor</vm:dcterm>
                        </xs:appinfo>
                        <xs:documentation>
                            Entity responsible for contributions to the content of
                            the resource
                        </xs:documentation>
                    </xs:annotation>
                </xs:element>

                <xs:element name="date" type="vr:Date"
                            minOccurs="0" maxOccurs="unbounded">
                    <xs:annotation>
                        <xs:appinfo>
                            <vm:dcterm>Date</vm:dcterm>
                        </xs:appinfo>
                        <xs:documentation>
                            Date associated with an event in the life cycle of the
                            resource.
                        </xs:documentation>
                        <xs:documentation>
                            This will typically be associated with the creation or
                            availability (i.e., most recent release or version) of
                            the resource.  Use the role attribute to clarify.
                        </xs:documentation>
                    </xs:annotation>
                </xs:element>

                <xs:element name="version" type="xs:token" minOccurs="0">
                    <xs:annotation>
                        <xs:documentation>
                            Label associated with creation or availablity of a version of
                            a resource.
                        </xs:documentation>
                    </xs:annotation>
                </xs:element>

                <xs:element name="contact" type="vr:Contact" maxOccurs="unbounded">
                    <xs:annotation>
                        <xs:documentation>
                            Information that can be used for contacting someone with
                            regard to this resource.
                        </xs:documentation>
                    </xs:annotation>
                </xs:element>

            </xs:sequence>
        </xs:complexType>

        <xs:complexType name="ResourceName">
            <xs:annotation>
                <xs:documentation>
                    the name of a potentially registered resource.  That is, the entity
                    referred to may have an associated identifier.
                </xs:documentation>
            </xs:annotation>

            <xs:simpleContent>
                <xs:extension base="xs:token">

                    <xs:attribute name="ivo-id" type="vr:IdentifierURI">
                        <xs:annotation>
                            <xs:documentation>
                                The URI form of the IVOA identifier for the resource refered to
                            </xs:documentation>
                        </xs:annotation>
                    </xs:attribute>

                </xs:extension>
            </xs:simpleContent>
        </xs:complexType>

        <xs:complexType name="Contact">
            <xs:annotation>
                <xs:documentation>
                    Information that can be used for contacting someone
                </xs:documentation>
            </xs:annotation>
            <xs:sequence>
                <xs:element name="name" type="vr:ResourceName">
                    <xs:annotation>
                        <xs:documentation>
                            the name or title of the contact person.
                        </xs:documentation>
                        <xs:documentation>
                            This can be a person's name, e.g. "John P. Jones" or
                            a group, "Archive Support Team".
                        </xs:documentation>
                    </xs:annotation>
                </xs:element>

                <xs:element name="address" type="xs:token" minOccurs="0">
                    <xs:annotation>
                        <xs:documentation>the contact mailing address</xs:documentation>
                        <xs:documentation>
                            All components of the mailing address are given in one
```



```
                        string, e.g. "3700 San Martin Drive, Baltimore, MD 21218 USA".
                    </xs:documentation>
                </xs:annotation>
            </xs:element>

            <xs:element name="email" type="xs:token" minOccurs="0">
                <xs:annotation>
                    <xs:documentation>the contact email address</xs:documentation>
                </xs:annotation>
            </xs:element>

            <xs:element name="telephone" type="xs:token" minOccurs="0">
                <xs:annotation>
                    <xs:documentation>the contact telephone number</xs:documentation>
                    <xs:documentation>
                        Complete international dialing codes should be given, e.g.
                        "+1-410-338-1234".
                    </xs:documentation>
                </xs:annotation>
            </xs:element>

        </xs:sequence>
    </xs:complexType>

    <xs:complexType name="Creator">
        <xs:annotation>
            <xs:documentation>
                The entity (e.g. person or organisation) primarily responsible
                for creating something
            </xs:documentation>
        </xs:annotation>

        <xs:sequence>
            <xs:element name="name" type="vr:ResourceName">
                <xs:annotation>
                    <xs:documentation>
                        the name or title of the creating person or organization
                    </xs:documentation>
                    <xs:documentation>
                        Users of the creation should use this name in
                        subsequent credits and acknowledgements.
                    </xs:documentation>
                </xs:annotation>
            </xs:element>

            <xs:element name="logo" type="vr:PaddedURI" minOccurs="0">
                <xs:annotation>
                    <xs:documentation>
                        URL pointing to a graphical logo, which may be used to help
                        identify the information source
                    </xs:documentation>
                </xs:annotation>
            </xs:element>

        </xs:sequence>
    </xs:complexType>

    <xs:complexType name="Date">
        <xs:simpleContent>
            <xs:extension base="vr:UTCDateTime">
                <xs:attribute name="role" type="xs:string" default="representative">
                    <xs:annotation>
                        <xs:documentation>
                            A string indicating what the date refers to.
                        </xs:documentation>
                        <xs:documentation>
                            While this vocabulary is uncontrolled, recognized strings
                            include "creation", indicating the date that the resource
                            itself was created, and "update", indicating when the
                            resource was updated last.  The default value,
                            "representative", means that the date is a rough
                            representation of the time coverage of the resource.
                        </xs:documentation>
                        <xs:documentation>
                            Note that this date refers to the resource; dates describing
                            the metadata description of the resource are handled by
                            the "created" and "updated" attributes of the Resource
                            element.
                        </xs:documentation>
                    </xs:annotation>
                </xs:attribute>
            </xs:extension>
        </xs:simpleContent>
    </xs:complexType>

    <xs:complexType name="Content">
        <xs:annotation>
            <xs:documentation>
                Information regarding the general content of a resource
            </xs:documentation>
        </xs:annotation>

        <xs:sequence>
            <xs:element name="subject" type="xs:token" maxOccurs="unbounded">
                <xs:annotation>
                    <xs:appinfo>
                        <vm:dcterm>Subject</vm:dcterm>
                    </xs:appinfo>
                    <xs:documentation>
                        a topic, object type, or other descriptive keywords
                        about the resource.
                    </xs:documentation>
                    <xs:documentation>
```



```xml
                    Terms for Subject should be drawn from the IAU Astronomy
                    Thesaurus (http://msowww.anu.edu.au/library/thesaurus/).
                </xs:documentation>
            </xs:annotation>
        </xs:element>

        <xs:element name="description" type="xs:token">
            <xs:annotation>
                <xs:appinfo>
                    <vm:dcterm>Description</vm:dcterm>
                </xs:appinfo>
                <xs:documentation>
                    An account of the nature of the resource
                </xs:documentation>
                <xs:documentation>
                    The description may include but is not limited to an abstract,
                    table of contents, reference to a graphical representation of
                    content or a free-text account of the content.
                </xs:documentation>
            </xs:annotation>
        </xs:element>

        <xs:element name="source" type="vr:Source" minOccurs="0">
            <xs:annotation>
                <xs:appinfo>
                    <vm:dcterm>Source</vm:dcterm>
                </xs:appinfo>
                <xs:documentation>
                    a bibliographic reference from which the present resource is
                    derived or extracted.
                </xs:documentation>
                <xs:documentation>
                    This is intended to point to an article in the published
                    literature.  An ADS Bibcode is recommended as a value when
                    available.
                </xs:documentation>
            </xs:annotation>
        </xs:element>

        <xs:element name="referenceURL" type="vr:PaddedURI">
            <xs:annotation>
                <xs:documentation>
                    URL pointing to a human-readable document describing this
                    resource.
                </xs:documentation>
            </xs:annotation>
        </xs:element>

        <xs:element name="type" type="vr:Type"
                    minOccurs="0" maxOccurs="unbounded">
            <xs:annotation>
                <xs:appinfo>
                    <vm:dcterm>Type</vm:dcterm>
                </xs:appinfo>
                <xs:documentation>
                    Nature or genre of the content of the resource
                </xs:documentation>
            </xs:annotation>
        </xs:element>

        <xs:element name="contentLevel" type="vr:ContentLevel"
                    minOccurs="0" maxOccurs="unbounded">
            <xs:annotation>
                <xs:appinfo>
                    <vm:dcterm>Subject</vm:dcterm>
                    <vm:dcterm>Subject.ContentLevel</vm:dcterm>
                </xs:appinfo>
                <xs:documentation>
                    Description of the content level or intended audience
                </xs:documentation>
            </xs:annotation>
        </xs:element>

        <xs:element name="relationship" type="vr:Relationship"
                    minOccurs="0" maxOccurs="unbounded">
            <xs:annotation>
                <xs:documentation>
                    a description of a relationship to another resource.
                </xs:documentation>
                <xs:documentation>
                    Because this element's type is abstract, an xsi:type must be
                    to indicate the set of relationship types that are valid.
                </xs:documentation>
            </xs:annotation>
        </xs:element>

    </xs:sequence>

</xs:complexType>

<xs:complexType name="Source">
    <xs:simpleContent>
        <xs:extension base="xs:token">
            <xs:attribute name="format" type="xs:string">
                <xs:annotation>
                    <xs:documentation>
                        The reference format.  Recognized values include "bibcode",
                        referring to a standard astronomical bibcode
                        (http://cdsweb.u-strasbg.fr/simbad/refcode.html).
                    </xs:documentation>
                </xs:annotation>
            </xs:attribute>
        </xs:extension>
```



```xml
      </xs:simpleContent>
   </xs:complexType>

   <xs:simpleType name="Type">
      <xs:restriction base="xs:token">
         <xs:enumeration value="Other">
            <xs:annotation>
               <xs:documentation>
                  resource that does not fall into any of the category names
                  currently defined.
               </xs:documentation>
            </xs:annotation>
         </xs:enumeration>
         <xs:enumeration value="Archive">
            <xs:annotation>
               <xs:documentation>
                  Collection of pointed observations
               </xs:documentation>
            </xs:annotation>
         </xs:enumeration>
         <xs:enumeration value="Bibliography">
            <xs:annotation>
               <xs:documentation>
                  Collection of bibliographic reference, abstracts, and
                  publications
               </xs:documentation>
            </xs:annotation>
         </xs:enumeration>
         <xs:enumeration value="Catalog">
            <xs:annotation>
               <xs:documentation>
                  Collection of derived data, primarily in tabular form
               </xs:documentation>
            </xs:annotation>
         </xs:enumeration>
         <xs:enumeration value="Journal">
            <xs:annotation>
               <xs:documentation>
                  Collection of scholarly publications under common editorial
                  policy
               </xs:documentation>
            </xs:annotation>
         </xs:enumeration>
         <xs:enumeration value="Library">
            <xs:annotation>
               <xs:documentation>
                  Collection of published materials (journals, books, etc.)
               </xs:documentation>
            </xs:annotation>
         </xs:enumeration>
         <xs:enumeration value="Simulation">
            <xs:annotation>
               <xs:documentation>
                  Theoretical simulation or model
               </xs:documentation>
            </xs:annotation>
         </xs:enumeration>
         <xs:enumeration value="Survey">
            <xs:annotation>
               <xs:documentation>
                  Collection of observations covering substantial and
                  contiguous areas of the sky
               </xs:documentation>
            </xs:annotation>
         </xs:enumeration>
         <xs:enumeration value="Transformation">
            <xs:annotation>
               <xs:documentation>
                  A service that transforms data
               </xs:documentation>
            </xs:annotation>
         </xs:enumeration>
         <xs:enumeration value="Education">
            <xs:annotation>
               <xs:documentation>
                  Collection of materials appropriate for educational use, such
                  as teaching resources, curricula, etc.
               </xs:documentation>
            </xs:annotation>
         </xs:enumeration>
         <xs:enumeration value="Outreach">
            <xs:annotation>
               <xs:documentation>
                  Collection of materials appropriate for public outreach, such
                  as press releases and photo galleries
               </xs:documentation>
            </xs:annotation>
         </xs:enumeration>
         <xs:enumeration value="EPOResource">
            <xs:annotation>
               <xs:documentation>
                  Collection of materials that may be suitable for EPO
                  products but which are not in final product form, as in Type
                  Outreach or Type Education.  EPOResource would apply,
                  e.g., to archives with easily accessed preview images or to
                  surveys with easy-to-use images.
               </xs:documentation>
            </xs:annotation>
         </xs:enumeration>
         <xs:enumeration value="Animation">
            <xs:annotation>
               <xs:documentation>
                  Animation clips of astronomical phenomena
```



```
                    </xs:documentation>
                  </xs:annotation>
                </xs:enumeration>
                <xs:enumeration value="Artwork">
                  <xs:annotation>
                    <xs:documentation>
                      Artists' renderings of astronomical phenomena or objects
                    </xs:documentation>
                  </xs:annotation>
                </xs:enumeration>
                <xs:enumeration value="Background">
                  <xs:annotation>
                    <xs:documentation>
                      Background information on astronomical phenomena or objects
                    </xs:documentation>
                  </xs:annotation>
                </xs:enumeration>
                <xs:enumeration value="BasicData">
                  <xs:annotation>
                    <xs:documentation>
                      Compilations of basic astronomical facts about objects,
                      such as approximate distance or membership in constellation.
                    </xs:documentation>
                  </xs:annotation>
                </xs:enumeration>
                <xs:enumeration value="Historical">
                  <xs:annotation>
                    <xs:documentation>
                      Historical information about astronomical objects
                    </xs:documentation>
                  </xs:annotation>
                </xs:enumeration>
                <xs:enumeration value="Photographic">
                  <xs:annotation>
                    <xs:documentation>
                      Publication-quality photographs of astronomical objects
                    </xs:documentation>
                  </xs:annotation>
                </xs:enumeration>
                <xs:enumeration value="Press">
                  <xs:annotation>
                    <xs:documentation>
                      Press releases about astronomical objects
                    </xs:documentation>
                  </xs:annotation>
                </xs:enumeration>
                <xs:enumeration value="Organisation">
                  <xs:annotation>
                    <xs:documentation>
                      An organization that is a publisher or curator of other
                      resources.
                    </xs:documentation>
                  </xs:annotation>
                </xs:enumeration>
                <xs:enumeration value="Project">
                  <xs:annotation>
                    <xs:documentation>
                      A project that is a publisher or curator of other resources
                    </xs:documentation>
                  </xs:annotation>
                </xs:enumeration>
                <xs:enumeration value="Registry">
                  <xs:annotation>
                    <xs:documentation>
                      a query service for which response is a structured
                      description of resources.
                    </xs:documentation>
                  </xs:annotation>
                </xs:enumeration>
              </xs:restriction>
            </xs:simpleType>

            <xs:simpleType name="ContentLevel">
              <xs:restriction base="xs:token">
                <xs:enumeration value="General">
                  <xs:annotation>
                    <xs:documentation>
                      Resource provides information appropriate for all users
                    </xs:documentation>
                  </xs:annotation>
                </xs:enumeration>
                <xs:enumeration value="Elementary Education">
                  <xs:annotation>
                    <xs:documentation>
                      Resource provides information appropriate for use in elementary
                      education (e.g. approximate ages 6-11)
                    </xs:documentation>
                  </xs:annotation>
                </xs:enumeration>
                <xs:enumeration value="Middle School Education">
                  <xs:annotation>
                    <xs:documentation>
                      Resource provides information appropriate for use in middle
                      school education (e.g. approximate ages 11-14)
                    </xs:documentation>
                  </xs:annotation>
                </xs:enumeration>
                <xs:enumeration value="Secondary Education">
                  <xs:annotation>
                    <xs:documentation>
                      Resource provides information appropriate for use in elementary
                      education (e.g. approximate ages 14-18)
                    </xs:documentation>
```



```
          </xs:annotation>
        </xs:enumeration>
        <xs:enumeration value="Community College">
          <xs:annotation>
            <xs:documentation>
              Resource provides information appropriate for use in
              community/junior college or early university education.
            </xs:documentation>
          </xs:annotation>
        </xs:enumeration>
        <xs:enumeration value="University">
          <xs:annotation>
            <xs:documentation>
              Resource provides information appropriate for use in
              university education
            </xs:documentation>
          </xs:annotation>
        </xs:enumeration>
        <xs:enumeration value="Research">
          <xs:annotation>
            <xs:documentation>
              Resource provides information appropriate for
              supporting scientific research.
            </xs:documentation>
          </xs:annotation>
        </xs:enumeration>
        <xs:enumeration value="Amateur">
          <xs:annotation>
            <xs:documentation>
              Resource provides information of interest to
              amateur astronomers.
            </xs:documentation>
          </xs:annotation>
        </xs:enumeration>
        <xs:enumeration value="Informal Education">
          <xs:annotation>
            <xs:documentation>
              Resource provides information appropriate for education
              at museums, planetariums, and other centers of informal learning.
            </xs:documentation>
          </xs:annotation>
        </xs:enumeration>
      </xs:restriction>
    </xs:simpleType>

    <xs:complexType name="Relationship">
      <xs:annotation>
        <xs:documentation>
          A description of the relationship between one resource and one or
          more other resources.
        </xs:documentation>
      </xs:annotation>

      <xs:sequence>
        <xs:element name="relationshipType" type="xs:token">
          <xs:annotation>
            <xs:documentation>
              the named type of relationship
            </xs:documentation>
            <xs:documentation>
              The VOResource Core specification defines a standard
              set of names that are not enforced by this schema,
              but are otherwise required by the spec.
            </xs:documentation>
          </xs:annotation>
        </xs:element>

        <xs:element name="relatedResource" type="vr:ResourceName"
                    minOccurs="1" maxOccurs="unbounded">
          <xs:annotation>
            <xs:documentation>
              the name of resource that this resource is related to.
            </xs:documentation>
          </xs:annotation>
        </xs:element>
      </xs:sequence>
    </xs:complexType>

    <!--
      -  The Organisation resource type
      -->

    <xs:complexType name="Organisation">
      <xs:annotation>
        <xs:documentation>
          A named group of one or more persons brought together to pursue
          participation in VO applications.
        </xs:documentation>
        <xs:documentation>
          According to the Resource Metadata Recommendation, organisations
          "can be hierarchical and range in size and scope.  At a high level,
          an organisation could be a university, observatory, or government
          agency.  At a finer level, it could be a specific scientific
          project, mission, or individual researcher."
        </xs:documentation>
        <xs:documentation>
          The main purpose of an organisation as a registered resource is
          to serve as a publisher of other resources.
        </xs:documentation>
      </xs:annotation>
      <xs:complexContent>
        <xs:extension base="vr:Resource">
          <xs:sequence>
```



```xml
                    <xs:element name="facility" type="vr:ResourceName"
                                minOccurs="0" maxOccurs="unbounded">
                        <xs:annotation>
                            <xs:appinfo>
                                <vm:dcterm>Subject</vm:dcterm>
                            </xs:appinfo>
                            <xs:documentation>
                                the observatory or facility used to collect the data
                                contained or managed by this resource.
                            </xs:documentation>
                        </xs:annotation>
                    </xs:element>

                    <xs:element name="instrument" type="vr:ResourceName"
                                minOccurs="0" maxOccurs="unbounded">
                        <xs:annotation>
                            <xs:appinfo>
                                <vm:dcterm>Subject</vm:dcterm>
                                <vm:dcterm>Subject.Instrument</vm:dcterm>
                            </xs:appinfo>
                            <xs:documentation>
                                the Instrument used to collect the data contain or
                                managed by a resource.
                            </xs:documentation>
                        </xs:annotation>
                    </xs:element>

                </xs:sequence>
            </xs:extension>
        </xs:complexContent>
    </xs:complexType>

    <!--
       -  The Service resource type
       -->

    <xs:complexType name="Service">
        <xs:annotation>
            <xs:documentation>
                a resource that can be invoked by a client to perform some action
                on its behalf.
            </xs:documentation>
        </xs:annotation>
        <xs:complexContent>
            <xs:extension base="vr:Resource">
                <xs:sequence>

                    <xs:element name="rights" type="vr:Rights"
                                minOccurs="0" maxOccurs="unbounded">
                        <xs:annotation>
                            <xs:appinfo>
                                <vm:dcterm>Rights</vm:dcterm>
                            </xs:appinfo>
                            <xs:documentation>
                                Information about rights held in and over the resource.
                            </xs:documentation>
                            <xs:documentation>
                                This should be repeated for all Rights values that apply.
                            </xs:documentation>
                        </xs:annotation>
                    </xs:element>

                    <xs:element name="capability" type="vr:Capability"
                                minOccurs="0" maxOccurs="unbounded">
                        <xs:annotation>
                            <xs:documentation>
                                a description of a general capability of the
                                service and how to use it.
                            </xs:documentation>
                            <xs:documentation>
                                This describes a general function of the
                                service, usually in terms of a standard
                                service protocol (e.g. SIA), but not
                                necessarily.
                            </xs:documentation>
                            <xs:documentation>
                                A service can have many capabilities
                                associated with it, each reflecting different
                                aspects of the functionality it provides.
                            </xs:documentation>
                        </xs:annotation>
                    </xs:element>
                </xs:sequence>
            </xs:extension>
        </xs:complexContent>
    </xs:complexType>

    <xs:simpleType name="Rights">
        <xs:restriction base="xs:token">
            <xs:enumeration value="public">
                <xs:annotation>
                    <xs:documentation>
                        unrestricted, public access is allowed without
                        authentication.
                    </xs:documentation>
                </xs:annotation>
            </xs:enumeration>
            <xs:enumeration value="secure">
                <xs:annotation>
                    <xs:documentation>
                        authenticated, public access is allowed.
                    </xs:documentation>
```



```
                    </xs:annotation>
                </xs:enumeration>
                <xs:enumeration value="proprietary">
                    <xs:annotation>
                        <xs:documentation>
                            only proprietary access is allowed with authentication.
                        </xs:documentation>
                    </xs:annotation>
                </xs:enumeration>
            </xs:restriction>
        </xs:simpleType>

        <xs:complexType name="Capability">
            <xs:annotation>
                <xs:documentation>
                    a description of what the service does (in terms of
                    context-specific behavior), and how to use it (in terms of
                    an interface)
                </xs:documentation>
            </xs:annotation>

            <xs:sequence>
                <xs:element name="validationLevel" type="vr:Validation"
                            minOccurs="0" maxOccurs="unbounded">
                    <xs:annotation>
                        <xs:documentation>
                            A numeric grade describing the quality of the
                            capability description and interface, when applicable,
                            to be used to indicate the confidence an end-user
                            can put in the resource as part of a VO application
                            or research study.
                        </xs:documentation>
                        <xs:documentation>
                            See vr:ValidationLevel for an explanation of the
                            allowed levels.
                        </xs:documentation>
                    </xs:annotation>
                </xs:element>

                <xs:element name="description" type="xs:token" minOccurs="0">
                    <xs:annotation>
                        <xs:documentation>
                            A human-readable description of what this capability
                            provides as part of the over-all service
                        </xs:documentation>
                        <xs:documentation>
                            Use of this optional element is especially encouraged when
                            this capability is non-standard and is one of several
                            capabilities listed.
                        </xs:documentation>
                    </xs:annotation>
                </xs:element>

                <xs:element name="interface" type="vr:Interface"
                            minOccurs="0" maxOccurs="unbounded">
                    <xs:annotation>
                        <xs:documentation>
                            a description of how to call the service to access
                            this capability
                        </xs:documentation>
                        <xs:documentation>
                            Since the Interface type is abstract, one must describe
                            the interface using a subclass of Interface, denoting
                            it via xsi:type.
                        </xs:documentation>
                        <xs:documentation>
                            Multiple occurances can describe different interfaces to
                            the logically same capability--i.e. data or functionality.
                            That is, the inputs accepted and the output provides should
                            be logically the same.  For example, a WebBrowser interface
                            given in addition to a WebService interface would simply
                            provide an interactive, human-targeted interface to the
                            underlying WebService interface.
                        </xs:documentation>
                    </xs:annotation>
                </xs:element>
            </xs:sequence>

            <xs:attribute name="standardID" type="xs:anyURI">
                <xs:annotation>
                    <xs:documentation>
                        A URI identifier for a standard service.
                    </xs:documentation>
                    <xs:documentation>
                        This provides a unique way to refer to a service
                        specification standard, such as a Simple Image Access service.
                        The use of an IVOA identifier here implies that a
                        VOResource description of the standard is registered and
                        accessible.
                    </xs:documentation>
                </xs:annotation>
            </xs:attribute>
        </xs:complexType>

        <xs:complexType name="Interface" abstract="true">
            <xs:annotation>
                <xs:documentation>
                    A description of a service interface.
                </xs:documentation>
                <xs:documentation>
                    Since this type is abstract, one must use an Interface subclass
                    to describe an actual interface.
                </xs:documentation>
```



```
                <xs:documentation>
                    Additional interface subtypes (beyond WebService and WebBrowser) are
                    defined in the VODataService schema.
                </xs:documentation>
            </xs:annotation>
        </xs:sequence>

        <xs:sequence>
            <xs:element name="accessURL" type="vr:AccessURL"
                        minOccurs="1" maxOccurs="unbounded">
                <xs:annotation>
                    <xs:documentation>
                        The URL (or base URL) that a client uses to access the
                        service.  How this URL is to be interpreted and used
                        depends on the specific Interface subclass
                    </xs:documentation>
                    <xs:documentation>
                        When more than one URL is given, each represents an
                        alternative (i.e. mirror) endpoint whose behavior is
                        identical to all the other accessURLs listed.
                    </xs:documentation>
                    <xs:documentation>
                        Editor's note: this element assumes that
                        all registered services are inherently web based.
                    </xs:documentation>
                </xs:annotation>
            </xs:element>

            <xs:element name="securityMethod" type="vr:SecurityMethod"
                        minOccurs="0" maxOccurs="unbounded">
                <xs:annotation>
                    <xs:documentation>
                        the mechanism the client must employ to gain secure
                        access to the service.
                    </xs:documentation>
                    <xs:documentation>
                        when more than one method is listed, each one must
                        be employed to gain access.
                    </xs:documentation>
                </xs:annotation>
            </xs:element>

        </xs:sequence>

        <xs:attribute name="version" type="xs:string" default="1.0">
            <xs:annotation>
                <xs:documentation>
                    The version of a standard interface specification that this
                    interface complies with.  When the interface is
                    provided in the context of a Capability element, then
                    the standard being refered to is the one identified by
                    the Capability's standardID element.  If the standardID
                    is not provided, the meaning of this attribute is
                    undefined.
                </xs:documentation>
            </xs:annotation>
        </xs:attribute>

        <xs:attribute name="role" type="xs:NMTOKEN">
            <xs:annotation>
                <xs:documentation>
                    A tag name the identifies the role the interface plays
                    in the particular capability.  If the value is equal to
                    "std" or begins with "std:", then the interface refers
                    to a standard interface defined by the standard
                    referred to by the capability's standardID attribute.
                </xs:documentation>
                <xs:documentation>
                    For an interface complying with some registered
                    standard (i.e. has a legal standardID), the role can be
                    match against interface roles enumerated in standard
                    resource record.  The interface descriptions in
                    the standard record can provide default descriptions
                    so that such details need not be repeated here.
                </xs:documentation>
            </xs:annotation>
        </xs:attribute>
    </xs:complexType>

    <xs:complexType name="AccessURL">
        <xs:simpleContent>
            <xs:extension base="xs:anyURI">
                <xs:attribute name="use">
                    <xs:annotation>
                        <xs:documentation>
                            A flag indicating whether this should be interpreted as a base
                            URL, a full URL, or a URL to a directory that will produce a
                            listing of files.
                        </xs:documentation>
                        <xs:documentation>
                            The default value assumed when one is not given depends on the
                            context.
                        </xs:documentation>
                    </xs:annotation>
                    <xs:simpleType>
                        <xs:restriction base="xs:string">
                            <xs:whiteSpace value="collapse"/>
                            <xs:enumeration value="full">
                                <xs:annotation>
                                    <xs:documentation>
                                        Assume a full URL--that is, one that can be invoked
                                        directly without alteration.  This usually returns a
                                        single document or file.
                                    </xs:documentation>
```



```
                </xs:annotation>
              </xs:enumeration>
              <xs:enumeration value="base">
                <xs:annotation>
                  <xs:documentation>
                    Assume a base URL--that is, one requiring an extra portion
                    to be appended before being invoked.
                  </xs:documentation>
                </xs:annotation>
              </xs:enumeration>
              <xs:enumeration value="dir">
                <xs:annotation>
                  <xs:documentation>
                    Assume URL points to a directory that will return a listing
                    of files.
                  </xs:documentation>
                </xs:annotation>
              </xs:enumeration>
            </xs:restriction>
          </xs:simpleType>
        </xs:attribute>
      </xs:extension>
    </xs:simpleContent>
  </xs:complexType>

  <xs:complexType name="SecurityMethod">
    <xs:annotation>
      <xs:documentation>
        a description of a security mechanism.
      </xs:documentation>
      <xs:documentation>
        this type only allows one to refer to the mechanism via a
        URI.  Derived types would allow for more metadata.
      </xs:documentation>
    </xs:annotation>

    <xs:sequence/>

    <xs:attribute name="standardID" type="xs:anyURI">
      <xs:annotation>
        <xs:documentation>
          A URI identifier for a standard security mechanism.
        </xs:documentation>
        <xs:documentation>
          This provides a unique way to refer to a security
          specification standard.  The use of an IVOA identifier here
          implies that a VOResource description of the standard is
          registered and accessible.
        </xs:documentation>
      </xs:annotation>
    </xs:attribute>

  </xs:complexType>

  <xs:complexType name="WebBrowser">
    <xs:annotation>
      <xs:documentation>
        A (form-based) interface intended to be accesed interactively
        by a user via a web browser.
      </xs:documentation>
      <xs:documentation>
        The accessURL represents the URL of the web form itself.
      </xs:documentation>
    </xs:annotation>

    <xs:complexContent>
      <xs:extension base="vr:Interface">
        <xs:sequence/>
      </xs:extension>
    </xs:complexContent>
  </xs:complexType>

  <xs:complexType name="WebService">
    <xs:annotation>
      <xs:documentation>
        A Web Service that is describable by a WSDL document.
      </xs:documentation>
      <xs:documentation>
        The accessURL element gives the Web Service's endpoint URL.
      </xs:documentation>
    </xs:annotation>

    <xs:complexContent>
      <xs:extension base="vr:Interface">
        <xs:sequence>
          <xs:element name="wsdlURL" type="xs:anyURI"
                      minOccurs="0" maxOccurs="unbounded">
            <xs:annotation>
              <xs:documentation>
                The location of the WSDL that describes this
                Web Service.  If not provided, the location is
                assumed to be the accessURL with "?wsdl" appended.
              </xs:documentation>
              <xs:documentation>
                Multiple occurances should represent mirror copies of
                the same WSDL file.
              </xs:documentation>
            </xs:annotation>
          </xs:element>
        </xs:sequence>
      </xs:extension>
    </xs:complexContent>
  </xs:complexType>
```



```
</xs:schema>
```

## Appendix B: Change History

**Changes from `v1.02`:**

- converted to Recommendation

**Changes from `v1.01`:**

- `status` attribute is now required.
- added this Change History appendix.

**Changes from `v1.0`**

- dropped `PaddedString`, `PaddedURI` and replaced with `xs:token`.
- made `Validation`'s `validatedBy` attribute required.
- reference citation correction for SOAP, WSDL.